\documentclass[12pt,a4paper]{article}
\usepackage{amsfonts}
\usepackage{amssymb}
\usepackage{graphicx}
\usepackage{setspace}
\usepackage{natbib}
\usepackage{amsmath}
\usepackage{lscape}  
\usepackage{amsthm} 
\usepackage{afterpage} 
\usepackage{graphbox}
\usepackage{palatino}
\usepackage{paralist}
\usepackage{comment}
\usepackage{threeparttable}
\usepackage{caption}
\usepackage{subcaption}
\usepackage[dvipsnames]{xcolor}
\usepackage{multirow}
\usepackage{booktabs}
\usepackage{dsfont}
\usepackage{soul}
\usepackage{indentfirst}
\usepackage{enumerate}
\usepackage{longtable}
\usepackage{caption,bbm}
\usepackage{algpseudocode} 
\usepackage{tikz}
\usepackage{lscape}
\usepackage{rotating}
\usepackage{floatpag}
\floatpagestyle{empty} 
\usepackage{setspace}

\usetikzlibrary{arrows}
\usetikzlibrary{positioning}
\usetikzlibrary{calc}
\usetikzlibrary{positioning,arrows.meta}
\newdimen\nodeDist
\usetikzlibrary{decorations.shapes}
\tikzset{decorate sep/.style 2 args=
  {decorate,decoration={shape backgrounds,shape=circle,shape size=#1,shape sep=#2}}}

\tikzset{
  treenode/.style = {shape=rectangle, 
    draw, align=center,
    top color=white, bottom color=white},
  env/.style      = {treenode}
}
\tikzstyle{level 1}=[level distance=1.5cm, sibling distance=4cm, font=\normalsize]
\tikzstyle{level 2}=[level distance=1.5cm, sibling distance=2cm, font=\normalsize]
\tikzstyle{level 3}=[level distance=1.5cm, sibling distance=1cm, font=\normalsize]
\tikzstyle{level 4}=[level distance=1.5cm, sibling distance=.5cm, font=\normalsize]

\usepackage[hidelinks]{hyperref}

\usepackage{algorithm,algpseudocode}

\makeatletter
\renewcommand{\fnum@algorithm}{\fname@algorithm}
\makeatother

\usepackage{titlesec}
\titleformat*{\section}{\large\bfseries}
\titleformat*{\subsection}{\normalsize\bfseries}
\titleformat*{\subsubsection}{\normalsize\bfseries}
\titleformat*{\paragraph}{\normalsize\bfseries}

\titlespacing{\section}{0pt}{*1.5}{*1.5}
\titlespacing{\subsection}{0pt}{*1.5}{*1.5}
\titlespacing{\subsubsection}{0pt}{*1.2}{*1.2}
\titlespacing{\paragraph}{0pt}{*1.5}{*1.5}

\newcommand\variablename[1]{\mathop{\mathit{#1}}\nolimits}

\newcommand{\I}{\mathbf{I}}

\newcommand{\Z}{\mathbf{Z}}

\newcommand{\F}{\mathbf{F}}
\newcommand{\Y}{\mathbf{Y}}

\newcommand{\B}{\mathbf{B}}
\newcommand{\A}{\mathbf{A}}

\newcommand{\M}{\mathbf{M}}
\newcommand{\m}{\mathbf{m}}
\newcommand{\HH}{\mathbf{H}}

\newcommand{\D}{\mathbf{D}}

\newcommand{\Q}{\mathbf{Q}}
\newcommand{\f}{\mathbf{f}}
\newcommand{\z}{\mathbf{z}}
\newcommand{\ZZ}{\mathbf{Z}}
\newcommand{\K}{\mathbf{K}}
\newcommand{\G}{\mathbf{G}}
\newcommand{\abf}{\mathbf{a}}
\newcommand{\U}{\mathbf{U}}
\newcommand{\DD}{\mathbf{D}}
\newcommand{\lbf}{\mathbf{l}}

\newcommand{\bmu}{\boldsymbol{\mu}}

\newcommand{\y}{\mathbf{y}}

\usepackage[toc,page]{appendix}
\usepackage{setspace}
\usepackage[left = 1in, right = 1in, top = 1in, bottom = 1in]{geometry}
\usepackage{titlesec}
\titleformat*{\section}{\large\bfseries}
\titleformat*{\subsection}{\normalsize\bfseries}
\titleformat*{\subsubsection}{\normalsize\bfseries}
\titleformat*{\paragraph}{\normalsize\bfseries}

\titlespacing{\section}{0pt}{*1.5}{*1.5}
\titlespacing{\subsection}{0pt}{*1.5}{*1.5}
\titlespacing{\subsubsection}{0pt}{*1.2}{*1.2}
\titlespacing{\paragraph}{0pt}{*1.5}{*1.5}

\allowdisplaybreaks

\title{\vspace{-8mm}  \LARGE Machine Learning and the  Yield Curve:  \vspace{-2mm}   \\Tree-Based Macroeconomic Regime Switching}

\author{\large Siyu Bie \vspace{-3mm}   \\ East China Normal University\vspace{-3mm}\\City University of Hong Kong  \\  \large Jingyu He \vspace{-3mm}   \\ City University of Hong Kong \and 
\large Francis X. Diebold \vspace{-3mm}   \\ University of Pennsylvania \vspace{-3mm} \\ \\ \large Junye Li  \vspace{-3mm}  \\ Fudan University    \vspace{8mm}
}

\date{ First Draft: July 2024\\ \vspace{-2mm} This Draft: \today}

\begin{document}

	\maketitle
	
	\vspace{5mm}
	
	\begin{spacing}{1}
		
		\noindent \textbf{Abstract}: We explore tree-based macroeconomic regime-switching in the context of the dynamic Nelson-Siegel (DNS) yield-curve model. In particular, we customize the tree-growing algorithm to partition macroeconomic variables based on the DNS model's marginal likelihood, thereby identifying regime-shifting patterns in the yield curve. Compared to traditional Markov-switching models, our model offers clear economic interpretation via macroeconomic linkages and ensures computational simplicity. In an empirical application to U.S. Treasury yields, we find (1) important  yield-curve  regime switching, and (2) evidence that macroeconomic variables have predictive power for the yield curve when the federal funds rate is high, but not in other regimes, thereby refining the notion of yield-curve ``macro-spanning".

\bigskip 

\bigskip

\footnotesize{\noindent \textbf{Acknowledgments:} We gratefully acknowledge helpful input from Yacine Ait-Sahalia, Michael Bauer, Sid Chib, Nour Meddahi and Mike West, as well as participants at the Stevanovich Center Conference on Big Data and Artificial Intelligence in Econometrics, Finance, and Statistics (University of Chicago), 2025 NSF/CEME Conference in Bayesian Inference in Econometrics and Statistics, and the FinEML Conference (USI Lugano). We thank Jacob Broussard for research assistance.  All remaining errors are ours alone.}

%*****  We are also grateful to seminar participants at ***

\bigskip


\bigskip


\bigskip

\noindent \textbf{Contact Information}: siyubie2-c@my.cityu.edu.hk (Bie); fdiebold@sas.upenn.edu (Diebold); \\  jingyuhe@cityu.edu.hk (He);  li\_junye@fudan.edu.cn (Li) 

\end{spacing}

 \thispagestyle{empty}

%
%\clearpage
%\setstretch{1.5}
%\thispagestyle{empty} 
%\setcounter{tocdepth}{3}
%\tableofcontents
%\doublespacing
%

\clearpage
\setcounter{page}{1}

\thispagestyle{empty}

\section{Introduction}
\label{sec:intro}

Yield-curve modeling plays a crucial role in asset pricing, risk management, and monetary policy. The Nelson-Siegel yield-curve model \citep{nelson1987parsimonious} has long been popular among both researchers and practitioners due to its simultaneous accuracy and parsimony. \cite{diebold2006forecasting} introduce a \textit{dynamic} Nelson-Siegel (DNS) model, allowing for three unobserved pricing factors (level, slope, and curvature) that evolve  smoothly over time.\footnote{See \cite{diebold2013yield} for a full exposition with variations and extensions.} However, in practice, the yield curve does not always evolve smoothly. Its underlying factors can experience abrupt shifts, particularly in response to the macroeconomy. This highlights the importance of possible yield-curve regime switching, and the importance of understanding how yield-curve regimes relate to macroeconomic conditions.

Economists have studied yield-curve regime switching \citep[e.g., ][]{Dai2007, hevia2015estimating}, typically invoking Markov-switching behavior \`a la \cite{hamilton1989new} and \cite{kim1994dynamic}.\footnote{See also \cite{gray1996}, \cite{ang2002regime}, \cite{bansal2002term}, and \cite{xiang2013regime}.} However, the latent nature of the ``hidden Markov" regimes makes it challenging to understand what the regimes actually represent, resulting in a lack of clear economic interpretation. Although several studies attempt to interpret regimes in terms of macroeconomic conditions such as real activity or asset market volatility \citep[e.g.,][]{hamilton1989new, bansal2002term}, those interpretations often have the feel of ex post rationalization. In addition, \cite{chib2009analysis} propose a Bayesian affine yield curve model and \cite{chib2013change} study the change points of it.  Much remains to be done in terms of developing an economically interpretable methodology for detecting macroeconomics regime changes and relating them to the yield curve.

In this paper we fill the void by developing a novel DNS extension that incorporates regime switching, which we call \textit{macro-instrumented DNS regimes}. We make two related contributions. First, we allow the DNS yield factors to switch dynamics across macroeconomic regimes, which we detect using a customized tree structure based on the values of an observable set of macroeconomic variables.  Using Bayesian methods, we choose optimal split candidates based on the marginal DNS likelihood, which enhances regime interpretability and provides a macroeconomically meaningful understanding of yield-curve dynamics.\footnote{We therefore make contact with two important recent literatures -- one that uses goal-orientated tree-based clustering with economic targets \citep[e.g.,][]{cong2025growing, feng2024currency, cong2023sparse, patton2023generalized}, and one that implements Bayesian analysis of regression tree models \citep[e.g.,][]{chipman2010bart, he2019xbart, he2023stochastic, krantsevich2023stochastic}.}  
 
Second, we contribute to the long-standing debate on whether the yield curve spans the macroeconomy (``macro-spanning"), meaning that all current and past macroeconomic information of relevance for bond pricing is contained in the current yield curve. Such macro-spanning implies, among other things, that macroeconomic variables have no predictive content for the yield curve. The validity of macro-spanning, however, is far from uncontentious. On the one hand, macro-spanning is a key implication of the popular class of affine yield-curve models, as reviewed for example in \cite{bauer2017resolving}. In those models,

\begin{spacing}{1}
\begin{quote}
... the short-term interest rate is represented as an affine function of risk factors ... that include macroeconomic variables. Accordingly, the assumption of the absence of arbitrage and the usual form of the stochastic discount factor imply that model-implied yields are also affine in these risk factors. This linear mapping from macro factors to yields can ... be inverted to express the macro factors as a linear combination of yields. Hence, these models imply ... “spanning”...,

\quad \quad \quad \quad \quad \quad \quad \quad \quad \quad \quad \quad \quad \quad \quad \quad \quad \quad\quad  \citep{bauer2017resolving}
\end{quote}
\end{spacing}	

\noindent and it has found some empirical support, as in \cite{bauer2017resolving}. On the other hand, macro-spanning is rejected by several other studies, ranging  from the early work of \cite{ang2003no} through more recent work like \cite{joslin2014risk} and \cite{bekaert2021macro}.\footnote{See also \cite{dewachter2006macro},  \cite{diebold2006macroeconomy}, \cite{ludvigson2009macro}, \cite{duffee2011}, \cite{chernov2012term}, and \cite{coroneo2016unspanned}.} Those studies, however, are based on linear models, whereas we investigate macro-spanning through the more nuanced nonlinear lens of macro-instrumented regime switching, which allows for the possibility that  macro-spanning might hold in some macroeconomic regimes but not in others. 

We proceed as follows. In section \ref{sec:method} we review the DNS model and introduce our tree-based macro-instrumented regime-detection framework. In section \ref{sec:empirical} we present empirical results for U.S. treasury yields, and we conclude in section \ref{sec:conclusion}.

\section{Dynamic Nelson-Siegel with Macro-Instrumented Regime Switching}
\label{sec:method}
 
In this section we introduce our modeling framework and estimation strategy, and we examine its performance in a small simulation. 
 
 \subsection{The State-Space Model} \label{sec:DNSmodel}
   
   We first introduce the ``yields-only" dynamic Nelson-Siegel model \citep*{diebold2006forecasting} and its extension to the ``yields-macro" yield-curve model  \citep*{diebold2006macroeconomy} on which we will focus. 
   
   \subsubsection{Yields-Only}
   
   DNS extends the original static Nelson-Siegel model \citep{nelson1987parsimonious} by writing the time-$t$ maturity-$\tau$ yield $y_t(\tau)$ as:
\begin{equation}
	\begin{aligned}
		y_t(\tau) = L_{t} + S_{t} \left( \frac{1- e^{-\lambda \tau}}{\lambda \tau}\right) + C_{t}  \left( \frac{1- e^{-\lambda \tau}}{\lambda \tau} - e^{-\lambda \tau}\right)+\varepsilon_t(\tau),
	\end{aligned}
\end{equation}
$t=1, ..., T$, where the three parameters $L_t$, $S_t$, $C_t$ are allowed to vary over time and are interpreted as  level, slope, and curvature factors, respectively;  $\varepsilon_t(\tau)$ is a stochastic shock capturing pricing error; and $\lambda$ is a parameter. The  measurement equation relating the $N$ observed yields to the three latent factors is then
$$
	\begin{pmatrix}
		y_t(\tau_1) \\
		y_t(\tau_2) \\
		\cdots      \\
		y_t(\tau_N)
	\end{pmatrix} = \begin{pmatrix}
		1 & \frac{1- e^{-\lambda \tau_1}}{\lambda \tau_1} & \frac{1- e^{-\lambda \tau_1}}{\lambda \tau_1} - e^{-\lambda \tau_1} \\
		1 & \frac{1- e^{-\lambda \tau_2}}{\lambda \tau_2} & \frac{1- e^{-\lambda \tau_2}}{\lambda \tau_2} - e^{-\lambda \tau_2} \\
		\cdots                                                                                                                  \\
		1 & \frac{1- e^{-\lambda \tau_N}}{\lambda \tau_N} & \frac{1- e^{-\lambda \tau_N}}{\lambda \tau_N} - e^{-\lambda \tau_N}
	\end{pmatrix}
	\begin{pmatrix}
		L_t \\ S_t \\ C_t
	\end{pmatrix} +
	\begin{pmatrix}
		\varepsilon_t(\tau_1)\\
		\varepsilon_t(\tau_2) \\
		\cdots                \\
		\varepsilon_t(\tau_N)
	\end{pmatrix}.
$$
Finally, the DNS model assumes VAR(1) transition dynamics for the state vector $(L_t, S_t, C_t)'$, thus forming a well-defined state-space system.

We now allow for regime switching in the latent factors. Specifically, assume there are $G$ regimes in total, and let $z_{t} \in \{ 1, 2, \cdots, G \}$ indicate the time-$t$ regime. Conditional on regime, we write the model as
\begin{equation}\label{eqn:yieldonly2}
	\begin{aligned}
		\y_t & = \boldsymbol{\Lambda} \boldsymbol{\mu}_{z_t} + \boldsymbol{\Lambda} \F_t  + \boldsymbol{\varepsilon}_t, \\
		\mathbf{F}_t & = \boldsymbol{A}_{z_{t-1}}\F_{t-1} + \boldsymbol{\eta}_t,
	\end{aligned}
\end{equation} 
where $\f_t = (L_t, S_t, C_t)^T$ are the factors, $\mathbf{F}_t = \f_t - \boldsymbol{\mu}_{z_t}$ are the demeaned factors, and $\y_t = (y_t(\tau_1), \cdots, y_t(\tau_N))^T$ are the $N$ yields at the $t$-th period. In particular, we assume the transition equation may be affected by regime changes such that the mean of the factors, $\boldsymbol{\mu}_{z_{t-1}}$, and the VAR evolution matrix, $\mathbf{A}_{z_{t-1}}$, depend on last period's regime, $z_{t-1}$, which is determined conditional on the information set through period $t-1$.  Furthermore, we make the standard assumption that the measurement and transition disturbances are independent and jointly normal,
\begin{equation}\label{eqn:residual}
	\begin{pmatrix}
		\boldsymbol{\varepsilon}_t \\
		\boldsymbol{\eta}_t
	\end{pmatrix} \sim \mathcal{N}\left(\mathbf{0}, \begin{pmatrix}
		\mathbf{Q} & \mathbf{0}       \\
		\mathbf{0} & \mathbf{H}_{z_t}
	\end{pmatrix}\right).
\end{equation}
We allow the dense covariance matrix of the factor innovations, $\mathbf{H}_{z_t}$, to vary contemporaneously with $z_t$. In contrast, we assume sparse (diagonal) measurement-error covariance matrix, $\mathbf{Q}=\text{diag}(\sigma^2_1, \cdots, \sigma^2_N)$, so that the residuals of the yields at different maturities are uncorrelated.

\subsubsection{Yields-Macro}

The above yields-only model can be extended to a yields-macro version that allows exploration of  the relationship between the yield factors and macroeconomic variables. Following \cite{diebold2006macroeconomy}, we examine three major macroeconomic variables:  capacity utilization $\variablename{CU}_t$ as an indicator of real economic activity (``quantities"), the federal funds rate $\variablename{FFR}_t$ as an indicator of the central bank stance (``policy"), and inflation $\variablename{INFL}_t$ as an indicator of nominal activity (``prices"). We set $\m_t = (\variablename{CU}_t, \variablename{FFR}_t, \variablename{INFL}_t)^T$, and we use it to augment the 3-dimensional ``yields-only" model in Equation (\ref{eqn:yieldonly2}), producing a six-dimensional ``yields-macro" model state space model. 

Let $\M_t = \m_t-\boldsymbol{\mu}_{z_t}^m$ be the demeaned macroeconomic factors, let $\mathbf{FF}_t = (\F_t, \M_t)^T$, and assume that the yield and macroeconomic factors in $\mathbf{FF}_t$ follow undelated VAR(1) processes. Then the yields-macro augmented state-space model is
\begin{equation}\label{eqn:macro_yields}
	\begin{aligned}
		\begin{pmatrix}
			\y_t \\
			\m_t
		\end{pmatrix} & =\begin{pmatrix}
			                 \boldsymbol{\Lambda} & \mathbf{0} \\
			                 \mathbf{0}  & \mathbf{I_3}
		                 \end{pmatrix} \mathbf{FF}_t+\begin{pmatrix}
			                                    \boldsymbol{\Lambda} & \mathbf{0} \\
			                                    \mathbf{0}  & \mathbf{I_3}
		                                    \end{pmatrix}\boldsymbol{\mu}_{z_t}+\boldsymbol{\varepsilon}_t \\
		\mathbf{FF}_t            & =\mathbf{A}_{z_{t-1}} \mathbf{FF}_{t-1}+\boldsymbol{\eta}_t,
	\end{aligned}
\end{equation}
where $\mathbf{A}_{z_{t-1}} = \begin{pmatrix}\mathbf{A}_{z_{t-1}}^{FF} & \mathbf{A}_{z_{t-1}}^{FM} \\\A_{z_{t-1}}^{MF} & \A_{z_{t-1}}^{MM} \end{pmatrix}$, $\boldsymbol{\mu}_{z_t} = (\boldsymbol{\mu}_{z_t}^F, \boldsymbol{\mu}_{z_t}^m)^T$, $\boldsymbol{\varepsilon}_t = (\boldsymbol{\varepsilon}_{t}^y, \boldsymbol{\varepsilon}_{t}^m)^T$,  $\boldsymbol{\eta}_t = (\boldsymbol{\eta}_t^y, \boldsymbol{\eta}_t^m)^T$, and the measurement and transition shocks  are assumed to have the same stochastic structure as in Equation (\ref{eqn:residual}).  

In closing this section, we emphasize that we use ``yields-only", or ``yields-only DNS", or (e.g., if $G=3$) ``3-regime yields-only DNS" to refer to model (\ref{eqn:yieldonly2}), and we similarly use ``yields-macro", or ``yields-macro DNS", or ``3-regime yields-macro DNS" to refer to model (\ref{eqn:macro_yields}). In what follows we will focus almost exclusively on the yields-macro model (\ref{eqn:macro_yields}), which includes the yields-only model (\ref{eqn:yieldonly2}) as a special case and permits not only the incorporation of regime switching, but also exploration of interactions between yield and macro variables in various regimes.

\subsection{Bayesian Estimation} \label{estimation}

We proceed in two steps. First we treat estimation assuming exogenously-known regimes, and then we incorporate endogenous regime determination (learning).

\subsubsection{Regimes Known}  \label{known}

We take a full Bayesian approach, estimating all model parameters using the Kalman filter/smoother in conjunction with Markov chain Monte Carlo (MCMC). 

\paragraph{Priors.} We use standard conjugate priors for all parameters. 

For the diagonal elements of $\A_g$, which drive the factor autocorrelations, we use normal priors.  For the off-diagonal elements in $\A_g$ we use spike-and-slab priors \citep{george1993variable}, which shrink weak signals toward zero while maintaining strong signals at values close to OLS estimates, thereby helping with sparse matrix estimation. The spike-and-slab prior for $a_{j,k}^g$ (the $(j,k)$-th element of $\A_g$) is
\begin{equation*}
	\begin{aligned}
		\pi(a_{jj}^g)                   & \sim N(0,\xi_1^2), \quad \text{for diagonal elements}                                      \\
		\pi(a_{jk}^g\mid \gamma_{jk}^g) & \sim (1-\gamma_{jk}^g)N(0, \xi_0^2) + \gamma_{jk}^g N(0,\xi_1^2), \quad \text{for }j\neq k \\
		\gamma_{jk}^g                   & \sim \text{Bernoulli}(w).
	\end{aligned}
\end{equation*}
We set the hyperparameter $\omega$ to 0.5, corresponding to agnostic beliefs regarding the off-diagonal elements in $\A_g$. 

For the covariance matrices $\Q$ and $\HH_g$, we use inverse Wishart priors. Because $\Q$ is diagonal, the inverse Wishart prior degenerates to the inverse Gamma prior for each diagonal element $\sigma^2_i$. For the factor mean $\mu_g$ we use a normal prior, and for the decay parameter $\lambda$ we use a uniform prior. Thus we write
\begin{equation*}
		\HH_g   \sim IW(\m_0, \M_0),   \quad 	\sigma^2_i  \sim IG(\alpha, \beta),  \quad  \boldsymbol{\boldsymbol{\mu}}_g  \sim \mathcal{N}(\underline{\boldsymbol{\mu}}, \underline{\B}), \quad  \lambda  \sim \text{Unif}(a, b).
\end{equation*}
Note that, departing from \cite{diebold2006forecasting} where $\lambda$ is calibrated at a fixed constant (0.0609), we allow $\lambda$ to be learned from the data and updated by a random-walk Metropolis-Hastings step. See Appendix \ref{sec:gibbs} for details of all prior specifications and the Gibbs sampler for posterior inference.

\paragraph{Marginal Likelihood.} We now discuss the marginal likelihood of the yields-macro model.\footnote{The marginal likelihood of the yields-only model (\ref{eqn:yieldonly2}) is similar and is of course subsumed by the yields-macro model.} The marginal likelihood is central to Bayesian model comparison and selection, because unknown model parameters are ``integrated out", thereby accounting for parameter estimation uncertainty and making it solely a function of the data. Most importantly for our purposes, it will play a crucial role in facilitating endogenous regime determination, to which we will soon move in section \ref{sec:regime}.

Let $\boldsymbol{\Theta} = \left(\boldsymbol{\Lambda}, \F_t, \{\A_g\}_{g=1}^G, \{\boldsymbol{\mu}_g\}_{g=1}^G, \{\HH_{g}\}_{g=1}^G, \Q\right)$ denote all parameters and latent factors in the yields-macro model \eqref{eqn:macro_yields}, and let $L(\y_t, \m_t, \z_t \mid \boldsymbol{\Theta})$ be the corresponding likelihood function. The latent factors $\F_t$ are unknown parameters requiring estimation, and there can be as many as 1,800 for a sample size of 600 months with three latent factors. It is therefore helpful to first integrate out the latent factors $\F_t$, producing a partial marginal likelihood, which for one time period is 
\begin{equation*}
	\begin{aligned}
		L(\y_t, \m_t, &\z_t\mid \boldsymbol{\Lambda}, \{\A_g\}_{g=1}^G, \{\boldsymbol{\mu}_g\}_{g=1}^G, \{\HH_{g}\}_{g=1}^G, \Q) = \int L(\y_t, \m_t, \z_t \mid \Theta)d\F_t                     \\
		                            &= \mathcal{N}\left(\boldsymbol{\Lambda} \F_{t\mid t, \F_{t+1}} + \boldsymbol{\Lambda} \boldsymbol{\mu}_{z_t}, ~ \boldsymbol{\Lambda} \mathbf{P}_{t\mid t, \F_{t+1}} \boldsymbol{\Lambda}^T +\Q\right),
	\end{aligned}
\end{equation*}
 a normal density whose mean and covariance matrix we present in Appendix \ref{sec:gibbs}. The full partial marginal likelihood for all periods is then
\begin{equation}\label{eqn:likelihood1}
	\begin{aligned}
		L(\Y, \M, &\ZZ \mid  \boldsymbol{\Lambda}, \{\A_g\}_{g=1}^G, \{\boldsymbol{\mu}_g\}_{g=1}^G, \{\HH_g\}_{g=1}^G, \Q ) \\ &= \prod_{t=1}^T L(\y_t, \m_t, z_t\mid \boldsymbol{\Lambda}, \{\A_g\}_{g=1}^G, \{\boldsymbol{\mu}_g\}_{g=1}^G, \{\HH_g\}_{g=1}^G, \Q),
	\end{aligned}
\end{equation}
where $\Y = \{\y_t\}_{t=1}^T$, $\M = \{\m_t\}_{t=1}^T$ and $\ZZ = \{z_t\}_{t=1}^T$.

Now we proceed to integrate out the remaining parameters from the partial marginal likelihood (\ref{eqn:likelihood1}), yielding
\begin{equation}\label{eqn:likelihood}
		L(\Y, \M, \ZZ) = \int L(\Y, \M, \Z\mid \boldsymbol{\Lambda}, \{\A_g\}_{g=1}^G, \{\boldsymbol{\mu}_g\}_{g=1}^G, \{\HH_{g}\}_{g=1}^G, \Q ) d\boldsymbol{\Lambda} d\A_gd \boldsymbol{\mu}_g d\HH_g d\Q,
\end{equation} 
which lacks a closed-form expression.  To tackle this problem, we draw posterior samples using the Gibbs sampler, insert each pair of posterior draws into the partial marginal likelihood  (\ref{eqn:likelihood}), and then average across the posterior draws. This effectively integrates out the remaining parameters numerically.

\subsubsection{Regimes Estimated}\label{sec:regime}
 
Thus far we have assumed a known number of regimes $G$, as well as known period-by-period values of regimes $z_t\in\{1,2,\cdots, G\}$, $t=1, ..., T$. A realistic treatment, however, requires that regimes be learned from the data, jointly with all model parameters. In this section we present a  model-based and macro-instrumented clustering approach to detect regimes. Our approach is model-based since the clusters are chosen according to valuations of the marginal likelihood of the model in Equation (\ref{eqn:macro_yields}), and it is macro-instrumented since all clusters are defined explicitly according to values of a set of macroeconomic variables.

The Classification and Regression Tree (CART) of \cite{breiman1984classification} is one of the most successful machine learning non-parametric regression models for prediction. CART greedily and sequentially partitions data into multiple leaf nodes according to certain decision rules, and then fits a constant to each leaf node for local prediction, thereby approximating curves with locally-constant step functions.  We borrow this divide-and-conquer strategy from CART, partitioning the data based on the value of macroeconomic variables to detect regimes. However, a significant distinction lies in our approach's focus on fitting the yields-macro model in Equation (\ref{eqn:macro_yields}), where the choice of splitting values is determined by model fitness, specifically, the marginal likelihood of the model. Therefore, we interpret the decision tree more from the perspective of partitioning and identifying regimes rather than merely as a tool for generating step functions for prediction.

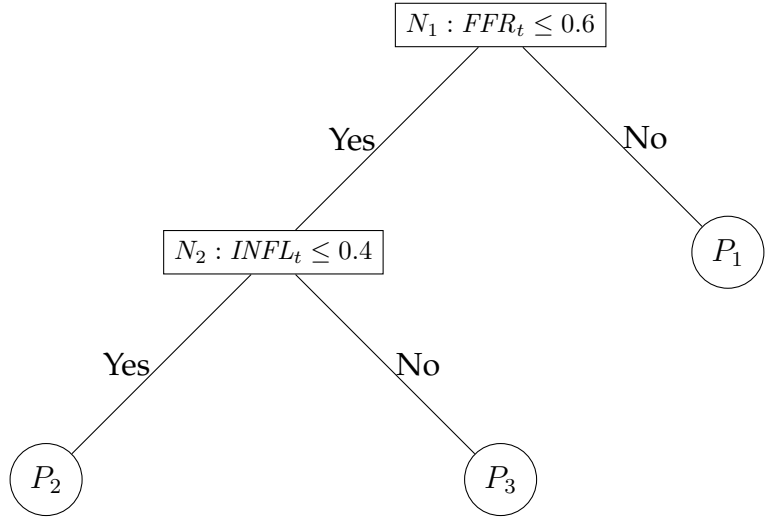
\begin{figure}[t]
	\caption{{An Illustrative Decision Tree}}	\label{fig:decision tree}
	
\begin{center}
		\begin{subfigure}{0.66\textwidth}
		\begin{center}
			\begin{tikzpicture}[
				scale=1.7,
				node/.style={%
					draw,
					rectangle,
				},
				node2/.style={%
					draw,
					circle,
				},
				]
				
				\node [node] (A) {\footnotesize{$N_1: \variablename{FFR}_t \leq 0.6$}};
				\path (A) ++(-135:\nodeDist) node [node] (B) {\footnotesize{$N_2: \variablename{INFL}_t \leq 0.4$}};
				\path (A) ++(-45:\nodeDist) node [node2] (C) {$P_1$};
				\path (B) ++(-135:\nodeDist) node [node2] (D) {$P_2$};
				\path (B) ++(-45:\nodeDist) node [node2] (E) {$P_3$};
				
				\draw (A) -- (B) node [left,pos=0.5] {Yes}(A);
				\draw (A) -- (C) node [right,pos=0.5] {No}(A);
				\draw (B) -- (D) node [left,pos=0.5] {Yes}(A);
				\draw (B) -- (E) node [right,pos=0.5] {No}(A);
			\end{tikzpicture}
		\end{center}
	\end{subfigure}
%	\vspace{0.2cm}

\end{center}
\begin{spacing}{1.0}  \noindent  \footnotesize 
	Notes: We illustrate a decision tree with two splits and three leaf nodes, creating three regimes based on macroeconomic variables, $\variablename{FFR}$ and $\variablename{INFL}$, at thresholds of 0.6 and 0.4, respectively.
\end{spacing}

\bigskip
\end{figure}

We present a simple illustrative decision-tree structure in Figure \ref{fig:decision tree}. It consists of two decision rules based on macroeconomic variables, $\variablename{FFR}_t \leq 0.6$ and $\variablename{INFL}_t$ $\leq 0.4$, representing the fed funds rate and inflation at thresholds of 0.6 and 0.4, respectively. The top node $N_1$ is the ``root node". Two split points partition the root node into three regimes, denoted $\{P_1, P_2, P_3\}$ and called ``leaf nodes" because  they have no further splits. For example, $P_3$ denotes a regime in which the fed funds rate is less than the 60\% historical quantile and inflation is higher than the 40\% historical quantile, and all the other regimes are interpreted similarly. For each time period $t$, starting from the top node of the decision tree, we compare the values of a set of macroeconomic variables at time $t$ with all decision rules and eventually navigate to one and only one leaf node. The corresponding regime label $z_{t}$ is simply the index of that leaf node.

Estimating the macro-instrumented regime-switching DNS model involves choosing the optimal splitting variables and thresholds from all candidates, along with estimating all parameters of the model in Equation (\ref{eqn:macro_yields}). Specifically, when determining the optimal splitting variables, we evaluate the marginal likelihood, as discussed in the previous section, where all other model parameters are integrated out \textit{a priori}.

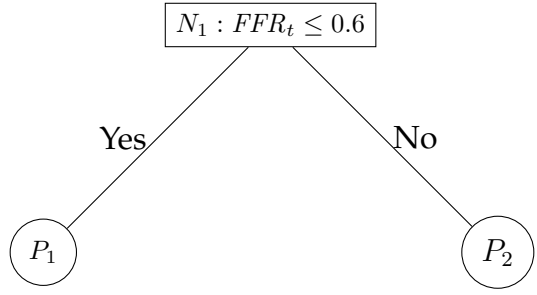
\begin{figure}[t]
	\caption{The First Decision-Tree Split}
	\label{fig:firstsplit}

\begin{center}
	
	\begin{subfigure}{0.66\textwidth}
		\begin{center}
			\begin{tikzpicture}[
				scale=1.7,
				node/.style={%
					draw,
					rectangle,
				},
				node2/.style={%
					draw,
					circle,
				},
				]
				
				\node [node] (A) {\footnotesize{$N_1: \variablename{FFR}_t \leq 0.6$}};
				\path (A) ++(-135:\nodeDist) node [node2] (B) {\footnotesize{$P_1$}};
				\path (A) ++(-45:\nodeDist) node [node2] (C) {$P_2$};
				\draw (A) -- (B) node [left,pos=0.5] {Yes}(A);
				\draw (A) -- (C) node [right,pos=0.5] {No}(A);
			\end{tikzpicture}
		\end{center}
	\end{subfigure}

\end{center}

\begin{spacing}{1.0}  \noindent  \footnotesize 
		
Notes: To determine the optimal first macro splitting variable and threshold, we evaluate several candidates ($FFR$, $INF$, $CU$) and thresholds (0.2, 0.4, 0.6, 0.8). The ``winning" candidate is $FFR$, with threshold $FFR \le 0.6$. 

\end{spacing}

\bigskip

\end{figure}

Now let us illustrate the splitting algorithm step by step. Before the first split, the root node itself is a leaf node. This means that all time periods are homogeneous, and only one regime encompasses all periods, i.e., $z_t = 1$ for all $t = 1,\cdots, T$. Then, we evaluate whether a split candidate is effective in partitioning the root, as illustrated in Figure \ref{fig:firstsplit}. 

Suppose that one split candidate based on the fed funds rate, $\mathcal{C}_i {:} \,\variablename{FFR}_t \leq 0.6$, partitions the data to two disjoint potential regimes $P_1$ and $P_2$, in which case we update the  regime indicators $z_t$ to reflect two regimes, $z_t^{\mathcal{C}_i} = 1$ if $\variablename{FFR}_t \leq 0.6$ and $z_t^{\mathcal{C}_i} = 2$ if $\variablename{FFR}_t > 0.6$. We denote the new set of regime indicators by $\ZZ^{\mathcal{C}_i} = \{z_t^{\mathcal{C}_i}\}_{t=1}^T$, where the superscript emphasizes its dependence on specific split candidate $\mathcal{C}_i$. 

Importantly, we use the joint marginal likelihood  to evaluate Equation (\ref{eqn:likelihood}) with the new regime indicators,
\begin{equation} \label{eqn:criterionfirstsplit}
	L(\mathcal{C}_i {:} \, \variablename{FFR}_t \leq 0.6) = L(\Y, \M, \ZZ^{\mathcal{C}_i}).
\end{equation}
We advocate using the marginal likelihood because it integrates out unknown parameters \textit{a priori}, accounting for parameter estimation uncertainty when determining regimes. While the yields $\Y$ and the macroeconomic variables $\M$ are fixed, the different split candidates create various regime partitions, yielding different regime indicators $\ZZ^{\mathcal{C}_i}$ and eventually leading to different evaluations of the split criterion in Equation (\ref{eqn:criterionfirstsplit}). We loop over all potential  macroeconomic split variables and thresholds and pick the first split point as the one with the highest joint marginal likelihood.

\begin{figure}[t]
	\caption{Candidates for the Second Decision-Tree Split}
	\label{fig:secondsplit}

\begin{center}
	
		\begin{subfigure}{0.48\textwidth}
		\begin{center}
			\begin{tikzpicture}[
				scale=1,
				node/.style={%
					draw,
					rectangle,
				},
				node2/.style={%
					draw,
					circle,
				},
				]
				
				\node [node] (A) {\footnotesize{$N_1: \variablename{FFR}_t \leq 0.6$}};
				\path (A) ++(-135:\nodeDist) node [node] (B) {\footnotesize{$N_2: \variablename{INFL}_t \leq 0.4$}};
				\path (A) ++(-45:\nodeDist) node [node2] (C) {$P_1$};
				\path (B) ++(-135:\nodeDist) node [node2] (D) {$P_2$};
				\path (B) ++(-45:\nodeDist) node [node2] (E) {$P_3$};
				
				\draw (A) -- (B) node [left,pos=0.5] {Yes}(A);
				\draw (A) -- (C) node [right,pos=0.5] {No}(A);
				\draw (B) -- (D) node [left,pos=0.5] {Yes}(A);
				\draw (B) -- (E) node [right,pos=0.5] {No}(A);
			\end{tikzpicture}
		\end{center}
		\caption{Splitting node \footnotesize{$N_2$ at $\variablename{INFL}_t \leq 0.4$}.}
	\end{subfigure}
	\begin{subfigure}{0.48\textwidth}
		\begin{center}
			\begin{tikzpicture}[
				scale=1,
				node/.style={%
					draw,
					rectangle,
				},
				node2/.style={%
					draw,
					circle,
				},
				] 
				
				\node [node] (A) {\footnotesize{$N_1: \variablename{FFR}_t \leq 0.6$}};
				\path (A) ++(-135:\nodeDist) node [node2] (B) {$P_1$};
				\path (A) ++(-45:\nodeDist) node [node] (C) {\footnotesize{$N_3: \variablename{CU} \leq 0.4$}};
				\path (C) ++(-135:\nodeDist) node [node2] (D) {$P_2$};
				\path (C) ++(-45:\nodeDist) node [node2] (E) {$P_3$};
				
				\draw (A) -- (B) node [left,pos=0.5] {Yes}(A);
				\draw (A) -- (C) node [right,pos=0.5] {No}(A);
				\draw (C) -- (D) node [left,pos=0.5] {Yes}(A);
				\draw (C) -- (E) node [right,pos=0.5] {No}(A);
			\end{tikzpicture}
		\end{center}
		\caption{Splitting node \footnotesize{$N_3$ at $\variablename{CU} \leq 0.4$}.}
	\end{subfigure}

\end{center}

\vspace{-2mm}

\begin{spacing}{1.0}  \noindent  \footnotesize 
	Notes: The left and right branches illustrate the potential candidates for the second split. Ultimately three leaf nodes (regimes) are available.
\end{spacing}

\bigskip

\end{figure}
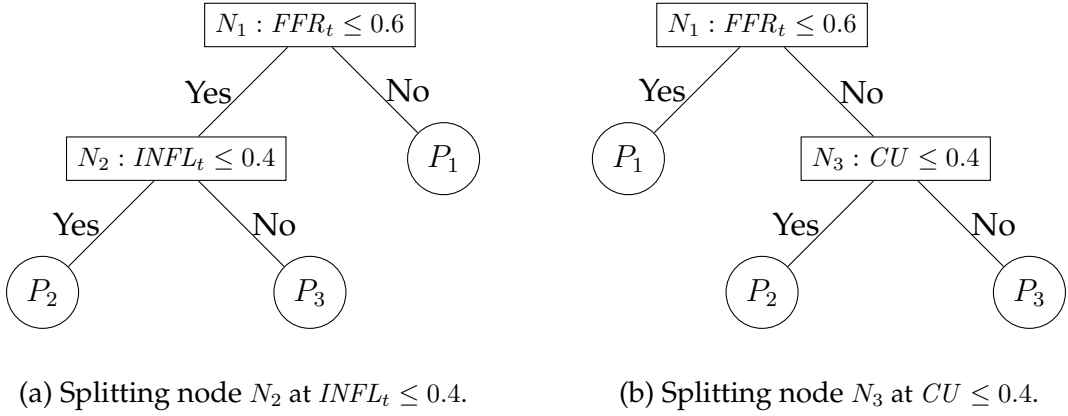

Once we determine the first split, we determine the second split in similar fashion, as illustrated in Figure \ref{fig:secondsplit}, where there are two potential ways to split, given the two leaf nodes created after the first step. The second split can happen at either leaf node created by the first split, and each has multiple potential macroeconomic variables and thresholds. Nevertheless, a candidate $\mathcal{C}_i$ will create one more regime, resulting in a total of three. Let $\Z^{\mathcal{C}_i}$ denote the new regime indicators following the candidate tree structure with three regimes. Thus, the split criterion is defined as $L(\Y, \M, \ZZ^{\mathcal{C}_i})$ again,
which varies with candidates since the indicators change while the yield and macro factors are fixed. We again pick the one with highest marginal likelihood.

All subsequent splits proceed similarly until a pre-specified stopping rule is triggered. In particular, we restrict the number of regimes to be no more than 3, consistent with the existing literature \citep[see, e.g.,][]{xiang2013regime}, and we also require the number of months in a regime to be at least 24 (2 years).

In Algorithm \ref{alg:split} we provide the pseudocode for determining the macro-instrumented regimes in the yields-macro model. Once the splitting algorithm stops, the regime labels $z_t$ for all time periods are determined explicitly by the tree structure, and we draw posterior samples of all parameters using the Gibbs sampler described in Appendix \ref{sec:gibbs}.

\begin{algorithm}[t]
	\small
	\caption{Macro-Instrumented Regime Clustering} 
\label{alg:split}
	\begin{algorithmic}[1]
		\State Search all current leaf nodes $\mathcal{P}= \{P_1, \cdots, P_J\}$.
		\For{each leaf node $P_i\in \mathcal{P}$}
		\If{$P_i$ satisfies minimal number of months in the node}
		\For{each split variable and threshold combination $\mathcal{C}_{i,s} {:} \, x_i \leq c_s $}.
		\State Partition $P_i$ to two leaf nodes $P_i^{\text{left}}$ and $P_i^{\text{right}}$.
		\State Update the regime indicators $\ZZ^{i,s}$ following the candidate structure
		\State Calculate the split criterion in Equation (\ref{eqn:criterionfirstsplit}) with new regime indicators
		$$
			L_{i,s} = L(\Y, \M, \ZZ^{i,s})
		$$
		\EndFor
		\EndIf
		\EndFor
		\State Pick the optimal split point $C_{i,s}$ with largest marginal likelihood $L_{i,s}$, partition leaf node $P_i$ according to $C_{i,s}$ and create two new leaf nodes.
		\State Check pre-specified stopping conditions. If satisfied then repeat from step 2, else stop.
	\end{algorithmic}

\end{algorithm}

\bigskip

\begin{table}[tp]
	\caption{Simulation Results for Parameter Estimation, Three-Regime Yields-Only Model}
	\label{tab:simulation}
	
	\vspace{-3mm}
	
	\begin{spacing}{1.0}
	\begin{center}
			
			\footnotesize
			\resizebox{\textwidth}{!}{
	\begin{tabular*}{\textwidth}{@{\extracolsep{\fill}}lccc|ccc|c}
		\toprule
		Regime 1                 & \multicolumn{3}{c}{$\A$} &    \multicolumn{3}{c}{$\HH$} &          $\boldsymbol{\mu}$ \\ \midrule
		& $L_{t-1}$                 & $S_{t-1}$ & $C_{t-1}$ & $L_{t}$                    & $S_{t}$ & $C_{t}$ &                    \\ \cmidrule{2-8} 
		\multirow{3}{*}{$L_{t}$} & 0.990     & 0.000     & 0.050     & 0.070   & -0.020  & -0.030  & 6.500              \\
		& 0.948     & 0.001     & 0.027     & 0.100   & -0.038  & -0.077  & 5.845              \\
		& (0.004)   & (0.000)   & (0.010)   & (0.006) & (0.006) & (0.023) & (0.087)            \\
		\\
		\multirow{3}{*}{$S_{t}$} & 0.000     & 0.980     & 0.100     & -0.020  & 0.050   & -0.070  & -1.800             \\
		& 0.010     & 0.973     & 0.081     & -0.038  & 0.084   & -0.027  & -1.627             \\
		& (0.006)   & (0.003)   & (0.008)   & (0.006) & (0.007) & (0.016) & (0.104)            \\
		\\
		\multirow{3}{*}{$C_{t}$} & 0.000     & 0.000     & 0.920     & -0.030  & -0.070  & 0.500   & -0.800             \\
		& -0.003    & 0.001     & 0.903     & -0.077  & -0.027  & 0.974   & -0.733             \\
		& (0.002)   & (0.000)   & (0.014)   & (0.023) & (0.016) & (0.104) & (0.119)            \\
		 \midrule
		Regime 2                 &                           &           &           &                            &         &         &                    \\ \midrule
		\multirow{3}{*}{$L_{t}$} & 0.980     & -0.040    & 0.000     & 0.100   & -0.080  & -0.050  & 6.000              \\
		& 0.980     & -0.001    & 0.000     & 0.124   & -0.097  & -0.114  & 5.517              \\
		& (0.001)   & (0.000)   & (0.000)   & (0.009) & (0.008) & (0.021) & (0.061)            \\
		\\
		\multirow{3}{*}{$S_{t}$} & 0.000     & 0.950     & 0.000     & -0.080  & 0.120   & 0.040   & -1.500             \\
		& 0.000     & 0.900     & 0.001     & -0.097  & 0.150   & 0.084   & -1.329             \\
		& (0.000)   & (0.004)   & (0.000)   & (0.008) & (0.009) & (0.024) & (0.077)            \\
		\\
		\multirow{3}{*}{$C_{t}$} & 0.000     & -0.200    & 0.900     & -0.050  & 0.040   & 0.900   & -0.500             \\
		& 0.002     & -0.279    & 0.877     & -0.114  & 0.084   & 1.057   & -0.659             \\
		& (0.001)   & (0.023)   & (0.009)   & (0.021) & (0.024) & (0.080) & (0.159)            \\
		 \midrule
		Regime 3                 &                           &           &           &                            &         &         &                    \\ \midrule
		\multirow{3}{*}{$L_{t}$} & 0.970     & -0.030    & 0.080     & 0.180   & -0.130  & -0.200  & 5.500              \\
		& 0.944     & -0.001    & 0.087     & 0.235   & -0.183  & -0.232  & 4.991              \\
		& (0.005)   & (0.001)   & (0.008)   & (0.012) & (0.013) & (0.034) & (0.068)            \\
		\\
		\multirow{3}{*}{$S_{t}$} & 0.000     & 0.920     & 0.000     & -0.130  & 0.250   & 0.200   & -1.200             \\
		& 0.006     & 0.920     & -0.010    & -0.183  & 0.268   & 0.215   & -0.971             \\
		& (0.004)   & (0.010)   & (0.006)   & (0.013) & (0.017) & (0.035) & (0.094)            \\
		\\
		\multirow{3}{*}{$C_{t}$} & 0.000     & 0.000     & 0.850     & -0.200  & 0.200   & 1.180   & -0.200             \\
		& -0.024    & 0.005     & 0.803     & -0.232  & 0.215   & 1.367   & -0.337             \\
		& (0.012)   & (0.004)   & (0.017)   & (0.034) & (0.035) & (0.119) & (0.114)            \\
		\bottomrule     
	\end{tabular*}
		}
		
	\end{center}
\end{spacing}
	
	\bigskip
	
	\begin{spacing}{1}
		
		\footnotesize 
		
		Notes: At each Monte Carlo replication we implement our Bayesian estimation procedure on simulated yield data, obtaining posterior mean parameter estimates. In each block of the table we show  the true parameter value  (first row),  the mean  of the posterior mean estimator across Monte Carlo replications  (second row), and the standard deviation  of the posterior mean estimator across Monte Carlo replications (third row, in parentheses), for each element of the transition matrix $\A$, the factor covariance matrix $\HH$, and the factor mean $\mu$.   See text for details.
	\end{spacing}
	
	\bigskip
	
\end{table}

\subsection{Simulation Evidence}\label{sec:simulation}

We now assess the efficacy of our Bayesian estimation procedure through a small simulation study based on the three-regime yields-only model in Equation (\ref{eqn:yieldonly2}). We use the yields-only model (\ref{eqn:yieldonly2}) -- despite the fact that we generally focus on the yields-macro model \eqref{eqn:macro_yields} in the rest of the paper -- because it is more tractable for simulation yet equally rich as a testbed for assessing our regime-switching detection and estimation procedure.

We calibrate the underlying true yields-only model parameters using empirical data.  We define the true regimes by two macroeconomic variables: the federal funds rate and inflation, covering the period from January 2001 to December 2022, encompassing a total of 264 months. To conduct the simulation, we utilize quantile values of these two variables within a rolling ten-year window. We categorize regimes as follows: in regime 1 the federal funds rate is greater than or equal to 0.6; in regime 2 both the federal funds rate and inflation are less than 0.6; and in regime 3  the federal funds rate is less than 0.6 but inflation is greater than or equal to 0.6. 

In keeping with our empirical analysis in the next section, we assume that there are 13 yield maturities: 3, 6, 9, 12, 24, 36, 48, 60, 72, 84, 96, 108, and 120 months. We generate monthly observations on yields with those maturities and perform 100 Bayesian estimation replications. In Table \ref{tab:simulation} we present simulation results for the transition matrix $\A$, factor covariance matrix $\HH$, and factor mean $\mu$. In each block of the table we report three things: the true parameter value,  the mean  of the posterior mean estimator across the 100 replications, and the standard deviation  of the posterior mean estimator across the 100 replications. For almost all parameters the biases are close to zero, and the standard deviations are small, suggesting that our approach provides accurate estimation of model parameters with endogenously-identified regimes.

\section{Empirical Analysis of U.S. Treasury Yields}\label{sec:empirical}

We now proceed to an empirical analysis of U.S. Treasury yields, examining regime switching in the yields-macro model \eqref{eqn:macro_yields}. Our primary interests are (1) determining and understanding the regime structure, and (2) characterizing  yield-factor and macro-factor interactions, if any, from the perspective of the macro-spanning debate. Most literature on yield-curve regimes considers only two possible regimes, which is potentially quite limiting, and moreover those regimes are latent rather than observed, which creates challenges for both estimation and interpretation.\footnote{See, for example,  \cite{Dai2007} and \cite{hevia2015estimating}.}  In contrast, we allow for three observable regimes, with clear macroeconomic interpretation. We choose splits sequentially, with the importance of each split decreasing with its order.  Hence the first two splits, which define the three regimes, are the most important.\footnote{Our small sample size -- just over 600 months of data, as we will soon discuss -- precludes allowing for more than three regimes, although one could of course do so in principle.}

\begin{table}[t]
	\caption{{Descriptive Statistics for  U.S. Yields}}\label{descriptive}
	
	\vspace{-3mm}
	
	\footnotesize

\begin{center}
	
	\begin{tabular*}{\textwidth}{@{\extracolsep{\fill}}ccccccccc}
		\toprule
		Maturity & \makebox[0.08\textwidth][c]{Mean} & \makebox[0.08\textwidth][c]{Std} & \makebox[0.08\textwidth][c]{Min} & \makebox[0.08\textwidth][c]{Max} & \makebox[0.08\textwidth][c]{$\widehat{\rho}(1)$} & \makebox[0.08\textwidth][c]{$\widehat{\rho}(6)$} & \makebox[0.08\textwidth][c]{$\widehat{\rho}(12)$} & \makebox[0.08\textwidth][c]{$\widehat{\rho}(30)$} \\ \midrule
		3        & 4.47                              & 3.51                             & 0.01                             & 15.95                            & 0.99                                             & 0.94                                             & 0.87                                              & 0.66                                              \\
		6        & 4.63                              & 3.56                             & 0.03                             & 16.13                            & 0.99                                             & 0.94                                             & 0.88                                              & 0.67                                              \\
		9        & 4.75                              & 3.57                             & 0.05                             & 16.11                            & 0.99                                             & 0.94                                             & 0.88                                              & 0.69                                              \\
		12       & 4.83                              & 3.57                             & 0.06                             & 15.96                            & 0.99                                             & 0.95                                             & 0.89                                              & 0.70                                              \\
		24       & 5.07                              & 3.53                             & 0.12                             & 15.72                            & 0.99                                             & 0.95                                             & 0.91                                              & 0.75                                              \\
		36       & 5.26                              & 3.46                             & 0.12                             & 15.57                            & 0.99                                             & 0.96                                             & 0.91                                              & 0.78                                              \\
		48       & 5.44                              & 3.39                             & 0.17                             & 15.48                            & 0.99                                             & 0.96                                             & 0.92                                              & 0.80                                              \\
		60       & 5.57                              & 3.31                             & 0.23                             & 15.20                            & 0.99                                             & 0.96                                             & 0.92                                              & 0.81                                              \\
		72       & 5.70                              & 3.26                             & 0.31                             & 14.99                            & 0.99                                             & 0.96                                             & 0.93                                              & 0.81                                              \\
		84       & 5.80                              & 3.20                             & 0.38                             & 14.95                            & 0.99                                             & 0.96                                             & 0.92                                              & 0.82                                              \\
		96       & 5.88                              & 3.15                             & 0.45                             & 14.94                            & 0.99                                             & 0.97                                             & 0.93                                              & 0.82                                              \\
		108      & 5.95                              & 3.11                             & 0.49                             & 14.95                            & 0.99                                             & 0.97                                             & 0.93                                              & 0.82                                              \\
		120      & 6.01                              & 3.04                             & 0.53                             & 14.94                            & 0.99                                             & 0.96                                             & 0.92                                              & 0.82                                              \\ \bottomrule
	\end{tabular*}

\end{center}

\begin{spacing}{1}

\footnotesize{Notes: We present descriptive statistics for U.S. Treasury bond yields at various maturities, measured in months. The last four columns are sample autocorrelations at displacements of  1, 6, 12 and 30 months, respectively. The sample period is August 1971 - December 2022. We express the mean (Mean), standard deviation (Std), minimum (Min) and maximum (Max) values as percentages.
}

\end{spacing}

\bigskip 

\end{table}

\subsection{Data}

We use the balanced zero-coupon U.S. Treasury  yield data constructed by \cite{liu2021reconstructing}, which employ a kernel-smoothing method with adaptive bandwidth selection.\footnote{The data are available at \url{https://sites.google.com/view/jingcynthiawu/yield-data}.} The data are monthly, from  August 1971 to December 2022, totaling 617 months, and they include 13 maturities of 3, 6, 9, 12, 24, 36, 48, 60, 72, 84, 96, 108, and 120 months.

\begin{figure}[t]
	%	\centering
	\caption{Time Series of U.S. Yields}
	\label{fig:yields_des}
	
	\begin{center}
		
		\vspace{-6mm}
		
		\includegraphics[width=0.9\textwidth,height=7cm]{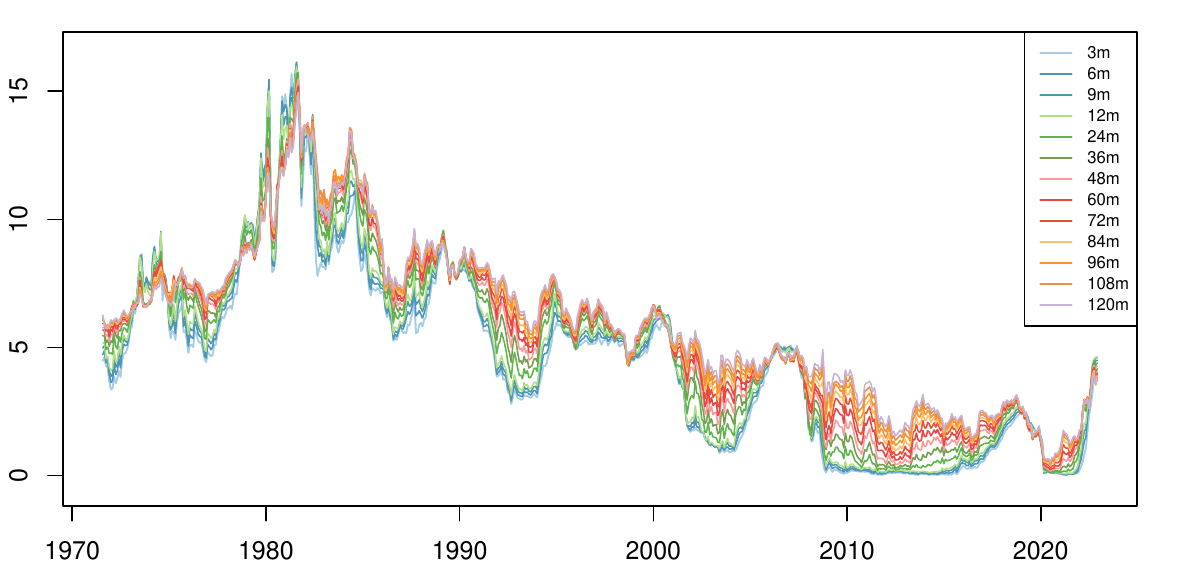}
		
	\end{center}

	\begin{spacing}{1.0}  \noindent  \footnotesize 
		Notes: We present time series of U.S. Treasury bond yields at maturities of 3, 6, 9, 12, 24, 36, 48, 60, 72, 84, 96, 108, and 120 months. The sample period is August 1971 - December 2022. 
	\end{spacing}
	
	\bigskip
	
\end{figure}

 We show yield data  descriptive statistics  in Table \ref{descriptive} and time series in Figure \ref{fig:yields_des}. Among other things, several well-known stylized facts are apparent: (1) yield curves generally slope upward (that is, the average yield generally increases with maturity) but are sometimes inverted; (2) yield volatility decreases with maturity; (3) long-maturity yields are more persistent than short-maturity yields.

We follow \cite{diebold2006macroeconomy} in using three macroeconomic factors in our yields-macro model: manufacturing capacity utilization ($\variablename{CU}_t$), the federal funds rate ($\variablename{FFR}_t$), and annual inflation ($\variablename{INFL}_t$).\footnote{All are from the Federal Reserve Economic Data (FRED) database of the Federal Reserve Bank of St. Louis, at \url{https://fred.stlouisfed.org/}.} Standardized versions of those variables also serve as the candidate split variables for classifying macroeconomic regimes. In particular, for regime classification we standardize the split variables  on a rolling window basis, replacing them with their quantiles based on historical ten-year rolling-windows, so that they lie in $[0, 1]$, and we use 0.2, 0.4, 0.6, and 0.8 as the candidate split thresholds. Our use of quantiles ensures that the splits are meaningful and comparable across space and time.\footnote{We emphasize that  yields-macro model estimation we leave the macroeconomic variables at their original, unstandardized, values, as appropriate. We convert to rolling quantiles  only for regime classification using algorithm \ref{alg:split}.}

\subsection{Yields-Macro Model Estimation}

\begin{figure}[t]
	\caption{Tree Structure and Time Series of Yield Factors, Three-Regime Yields-Macro Model}
	\label{fig:tree&factorsmodel2}

	%\bigskip

	\begin{subfigure}{0.4\textwidth}
		\begin{center}
			\begin{tikzpicture}[
					node/.style={%
							draw,
							rectangle,
						},
					node2/.style={%
							draw,
							circle,
						},
				]

				\node [node] (A) {\footnotesize{$N_1: \variablename{FFR}_t \leq 0.6$}};
				\path (A) ++(-135:\nodeDist) node [node] (B) {\footnotesize{$N_2: \variablename{INFL}_t \leq 0.6$}};
				\path (A) ++(-45:\nodeDist) node [node2] (C) {$P_1$};
				\path (B) ++(-135:\nodeDist) node [node2] (D) {$P_2$};
				\path (B) ++(-45:\nodeDist) node [node2] (E) {$P_3$};

				\draw (A) -- (B) node [left,pos=0.5] {Yes}(A);
				\draw (A) -- (C) node [right,pos=0.5] {No}(A);
				\draw (B) -- (D) node [left,pos=0.5] {Yes}(A);
				\draw (B) -- (E) node [right,pos=0.5] {No}(A);
			\end{tikzpicture}
			\caption{\footnotesize Tree Structure}
		\end{center}
	\end{subfigure}\hfill
	\begin{subfigure}{0.6\textwidth}
		\begin{center}
			\includegraphics[width=\textwidth]{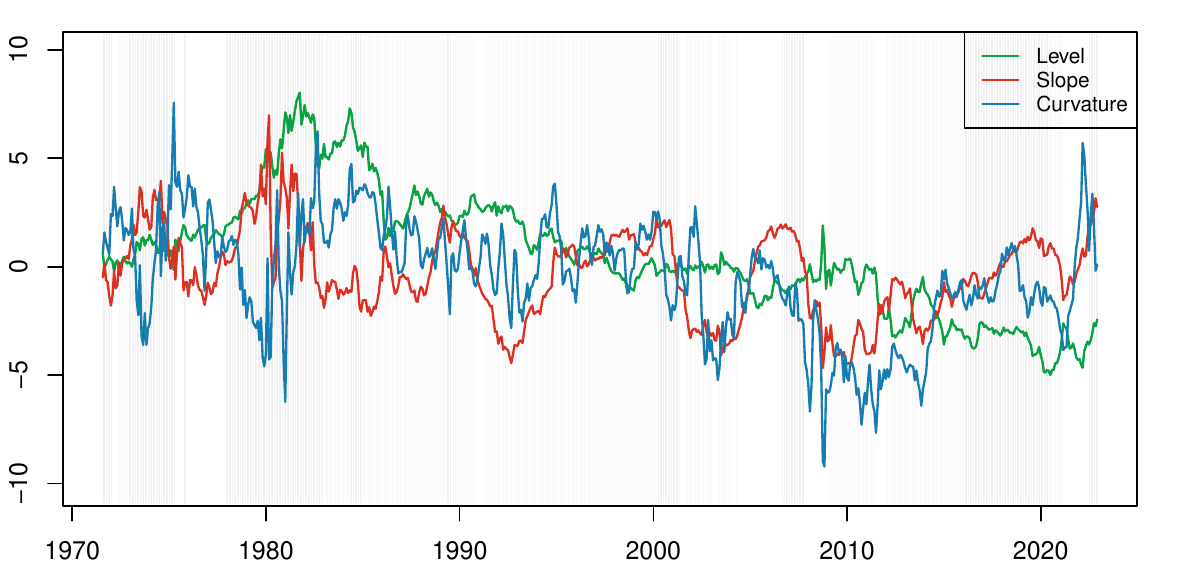}
			\caption{\footnotesize Yield Factors}
		\end{center}
	\end{subfigure}

\vspace{2mm}

\begin{spacing}{1.0}  \noindent  \footnotesize 
	Notes: In the left panel (a) we show the tree structure of estimated macro-instrumented regimes in the yields-macro model. It  partitions twice, based first on the federal funds rate and then on inflation, resulting in three regimes. In the right panel (b) we plot the three time series of latent yield factors extracted from the model, using background shading that indicates regime (dark gray for Regime 1, light gray for Regime 2, and white for Regime 3).
	
\end{spacing}

\bigskip

\end{figure}

In this section we present empirical results for the yields-macro regime-switching model in Equation (\ref{eqn:macro_yields}).  Our  first key result is that there is indeed evidence for regime switching, as evidenced by the three-regime model favored by the marginal likelihood.  As we show in the left panel of Figure \ref{fig:tree&factorsmodel2}, the first identified split is on $\variablename{FFR}_t$ (federal funds rate) at a threshold of 0.6, and the second is on $\variablename{INFL}_t$ (inflation), also at a threshold of 0.6. The two splits create three regimes, Regime 1 ($\variablename{FFR}_t$ $\geq$ 0.6) with 191 months, Regime 2 ($\variablename{FFR}_t$ $<$ 0.6 \& $\variablename{INFL}_t$ $<$ 0.6) with 310 months, and Regime 3 ($\variablename{FFR}_t$ $<$ 0.6 \& $\variablename{INFL}_t$ $\geq$ 0.6) with 116 months. In the right panel of Figure \ref{fig:tree&factorsmodel2} we present the three estimated time series of yield factors, with shaded background regions denoting the three identified regimes.

In Table \ref{tab:para_empirical_macroyields_3regime_case1} we present the posterior means and standard deviations of the model parameters for the three regimes, which brings us to our second key result: there is regime-specific  evidence both for and against macro-spanning. In Regimes 2 and 3, the estimated transition matrix $\A$ shows little or no effects of macro factors on yield factors, suggesting that the macro factors have little or no predictive value for the yield curve, so that macro spanning seems to be a reasonable approximation to the yields-macro dynamics in Regimes 2 and 3. In contrast, however, in Regime 1, which features high federal funds rates, we find a significant positive effect of inflation on the curvature factor, so that macro spanning is violated. 

\begin{table}[p]
	\caption{Parameter Estimates, Three-Regime Yields-Macro Model}
	\label{tab:para_empirical_macroyields_3regime_case1}
	
	\begin{center}
		
		\vspace{-6mm}
		
		\scriptsize
		\setlength{\tabcolsep}{1.1mm}{
			\begin{tabular*}{\textwidth}{@{\extracolsep{\fill}}lcccccc|cccccc|c}
				\toprule
				Regime 1                                 & \multicolumn{6}{c|}{$\A$}                                                                                                & \multicolumn{6}{c|}{$\HH$}                                                                             & $\boldsymbol{\mu}$ \\ \midrule
				& $L_{t-1}$ & $S_{t-1}$ & $C_{t-1}$ & $\variablename{CU}_{t-1}$ & $\variablename{FFR}_{t-1}$ & $\variablename{INFL}_{t-1}$ & $L_{t}$ & $S_{t}$ & $C_{t}$ & $\variablename{CU}_t$ & $\variablename{FFR}_t$ & $\variablename{INFL}_t$ &                    \\ \cmidrule{2-14} 
				\multirow{2}{*}{$L_{t}$}                 & 0.98*     & 0.02      & 0.00      & 0.00                      & 0.01                       & 0.00                        & 0.14*   & 0.00    & -0.24*  & 0.04                  & 0.08*                  & 0.02*                   & 5.95*              \\
				& (0.02)    & (0.03)    & (0.01)    & (0.00)                    & (0.02)                     & (0.01)                      & (0.02)  & (0.02)  & (0.04)  & (0.03)                & (0.02)                 & (0.01)                  & (0.10)             \\
				\multirow{2}{*}{$S_{t}$}                 & 0.00      & 0.92*     & 0.00      & 0.00                      & 0.00                       & 0.00                        &         & 0.65*   & -0.11   & 0.16*                 & 0.33*                  & 0.03                    & -1.46*             \\
				& (0.00)    & (0.03)    & (0.00)    & (0.00)                    & (0.00)                     & (0.01)                      &         & (0.07)  & (0.08)  & (0.07)                & (0.05)                 & (0.02)                  & (0.12)             \\
				\multirow{2}{*}{$C_{t}$}                 & 0.00      & -0.27*    & 0.78*     & 0.00                      & 0.01                       & 0.13*                       &         &         & 1.55*   & -0.06                 & -0.30*                 & -0.03                   & -0.12              \\
				& (0.03)    & (0.09)    & (0.04)    & (0.01)                    & (0.03)                     & (0.04)                      &         &         & (0.17)  & (0.12)                & (0.08)                 & (0.03)                  & (0.24)             \\
				\multirow{2}{*}{$\variablename{CU}_t$}   & 0.03      & 0.03      & 0.00      & 0.96*                     & -0.03                      & -0.04                       &         &         &         & 1.40*                 & 0.22*                  & 0.08*                   & 77.78*             \\
				& (0.09)    & (0.10)    & (0.02)    & (0.02)                    & (0.09)                     & (0.04)                      &         &         &         & (0.15)                & (0.07)                 & (0.03)                  & (0.21)             \\
				\multirow{2}{*}{$\variablename{FFR}_t$}  & 0.43*     & 0.55*     & 0.00      & 0.00                      & 0.59*                      & 0.00                        &         &         &         &                       & 0.56*                  & 0.04*                   & 4.53*              \\
				& (0.05)    & (0.05)    & (0.00)    & (0.00)                    & (0.04)                     & (0.00)                      &         &         &         &                       & (0.06)                 & (0.02)                  & (0.08)             \\
				\multirow{2}{*}{$\variablename{INFL}_t$} & 0.00      & 0.00      & -0.01     & 0.03*                     & 0.00                       & 0.98*                       &         &         &         &                       &                        & 0.08*                   & 3.45*              \\
				& (0.00)    & (0.00)    & (0.01)    & (0.00)                    & (0.00)                     & (0.01)                      &         &         &         &                       &                        & (0.01)                  & (0.07)             \\ \midrule
				Regime 2                                 &           &           &           &                           &                            &                             &         &         &         &                       &                        &                         &                    \\ \midrule
				\multirow{2}{*}{$L_{t}$}                 & 0.99*     & 0.00      & 0.00      & 0.00                      & 0.00                       & 0.00                        & 0.10*   & -0.08*  & -0.01   & 0.00                  & 0.00                   & 0.01*                   & 6.07*              \\
				& (0.01)    & (0.00)    & (0.00)    & (0.00)                    & (0.00)                     & (0.00)                      & (0.01)  & (0.01)  & (0.02)  & (0.01)                & (0.00)                 & (0.00)                  & (0.08)             \\
				\multirow{2}{*}{$S_{t}$}                 & 0.00      & 0.99*     & 0.00      & 0.00                      & 0.00                       & 0.00                        &         & 0.13*   & 0.02    & 0.02                  & 0.02*                  & -0.01                   & -1.91*             \\
				& (0.00)    & (0.01)    & (0.00)    & (0.00)                    & (0.00)                     & (0.00)                      &         & (0.01)  & (0.02)  & (0.01)                & (0.00)                 & (0.01)                  & (0.08)             \\
				\multirow{2}{*}{$C_{t}$}                 & 0.00      & 0.00      & 0.93*     & 0.00                      & 0.02                       & 0.00                        &         &         & 0.59*   & 0.03                  & 0.01                   & -0.01                   & -0.50              \\
				& (0.01)    & (0.02)    & (0.03)    & (0.00)                    & (0.03)                     & (0.01)                      &         &         & (0.06)  & (0.03)                & (0.01)                 & (0.01)                  & (0.25)             \\
				\multirow{2}{*}{$\variablename{CU}_t$}   & 0.02      & 0.00      & 0.01      & 0.97*                     & 0.00                       & 0.00                        &         &         &         & 0.32*                 & 0.01                   & 0.02*                   & 78.25*             \\
				& (0.02)    & (0.01)    & (0.02)    & (0.01)                    & (0.01)                     & (0.01)                      &         &         &         & (0.03)                & (0.01)                 & (0.01)                  & (0.15)             \\
				\multirow{2}{*}{$\variablename{FFR}_t$}  & 0.29*     & 0.28*     & 0.01      & 0.00                      & 0.73*                      & 0.00                        &         &         &         &                       & 0.03*                  & 0.00                    & 4.46*              \\
				& (0.03)    & (0.03)    & (0.01)    & (0.00)                    & (0.03)                     & (0.00)                      &         &         &         &                       & (0.00)                 & (0.00)                  & (0.03)             \\
				\multirow{2}{*}{$\variablename{INFL}_t$} & 0.00      & 0.00      & 0.00      & 0.00                      & 0.00                       & 0.99*                       &         &         &         &                       &                        & 0.06*                   & 3.64*              \\
				& (0.00)    & (0.00)    & (0.00)    & (0.00)                    & (0.00)                     & (0.01)                      &         &         &         &                       &                        & (0.00)                  & (0.04)             \\ \midrule
				Regime 3                                 &           &           &           &                           &                            &                             &         &         &         &                       &                        &                         &                    \\ \midrule
				\multirow{2}{*}{$L_{t}$}                 & 0.96*     & 0.00      & 0.05*     & 0.00                      & 0.01                       & 0.00                        & 0.16*   & -0.16*  & -0.31*  & -0.02                 & -0.02*                 & -0.01                   & 6.01*              \\
				& (0.02)    & (0.01)    & (0.01)    & (0.00)                    & (0.02)                     & (0.00)                      & (0.02)  & (0.03)  & (0.06)  & (0.03)                & (0.01*)                & (0.01)                  & (0.09)             \\
				\multirow{2}{*}{$S_{t}$}                 & 0.00      & 0.95*     & 0.00      & 0.00                      & 0.00                       & 0.00                        &         & 0.28*   & 0.32*   & 0.03                  & 0.04*                  & 0.02                    & -1.72*             \\
				& (0.01)    & (0.02)    & (0.00)    & (0.00)                    & (0.01)                     & (0.01)                      &         & (0.04)  & (0.07)  & (0.03)                & (0.01*)                & (0.02)                  & (0.11)             \\
				\multirow{2}{*}{$C_{t}$}                 & 0.00      & 0.00      & 0.92*     & 0.01                      & 0.01                       & 0.00                        &         &         & 1.30*   & 0.04                  & 0.03                   & 0.06                    & -0.43              \\
				& (0.03)    & (0.03)    & (0.03)    & (0.02)                    & (0.03)                     & (0.01)                      &         &         & (0.19)  & (0.08)                & (0.03)                 & (0.04)                  & (0.29)             \\
				\multirow{2}{*}{$\variablename{CU}_t$}   & 0.00      & 0.00      & 0.04      & 0.98*                     & 0.00                       & 0.05                        &         &         &         & 0.45*                 & 0.03*                  & 0.07*                   & 78.42*             \\
				& (0.02)    & (0.02)    & (0.04)    & (0.02)                    & (0.02)                     & (0.06)                      &         &         &         & (0.06)                & (0.02)                 & (0.03)                  & (0.15)             \\
				\multirow{2}{*}{$\variablename{FFR}_t$}  & 0.31*     & 0.32*     & 0.03*     & 0.01                      & 0.68*                      & 0.00                        &         &         &         &                       & 0.05*                  & 0.02*                   & 4.47*              \\
				& (0.06)    & (0.06)    & (0.01)    & (0.01)                    & (0.06)                     & (0.00)                      &         &         &         &                       & (0.01)                 & (0.01)                  & (0.05)             \\
				\multirow{2}{*}{$\variablename{INFL}_t$} & -0.02     & 0.00      & 0.00      & 0.02                      & 0.00                       & 0.98*                       &         &         &         &                       &                        & 0.13*                   & 3.76*              \\
				& (0.03)    & (0.01)    & (0.01)    & (0.02)                    & (0.02)                     & (0.01)                      &         &         &         &                       &                        & (0.02)                  & (0.05)             \\ \bottomrule
			\end{tabular*}
		}
		
	\end{center}
	
	\vspace{-3mm}
	
	\begin{spacing}{1} 	\scriptsize  
		Notes:  We show posterior means and standard deviations in each of the three regimes. Asterisks indicate that the posterior 95\% credible interval does not include 0. Because $\HH$ is a symmetric matrix, we show only its diagonal and upper-right elements.
	\end{spacing}
	
	\bigskip
	
\end{table}

\begin{figure}[tb]
	\caption{Actual and Fitted Yield Curves, Three-Regime  Yields-Macro  Model}
	
	\vspace{-5mm}
	
	\begin{center}
		
		\includegraphics[width=0.8\textwidth]{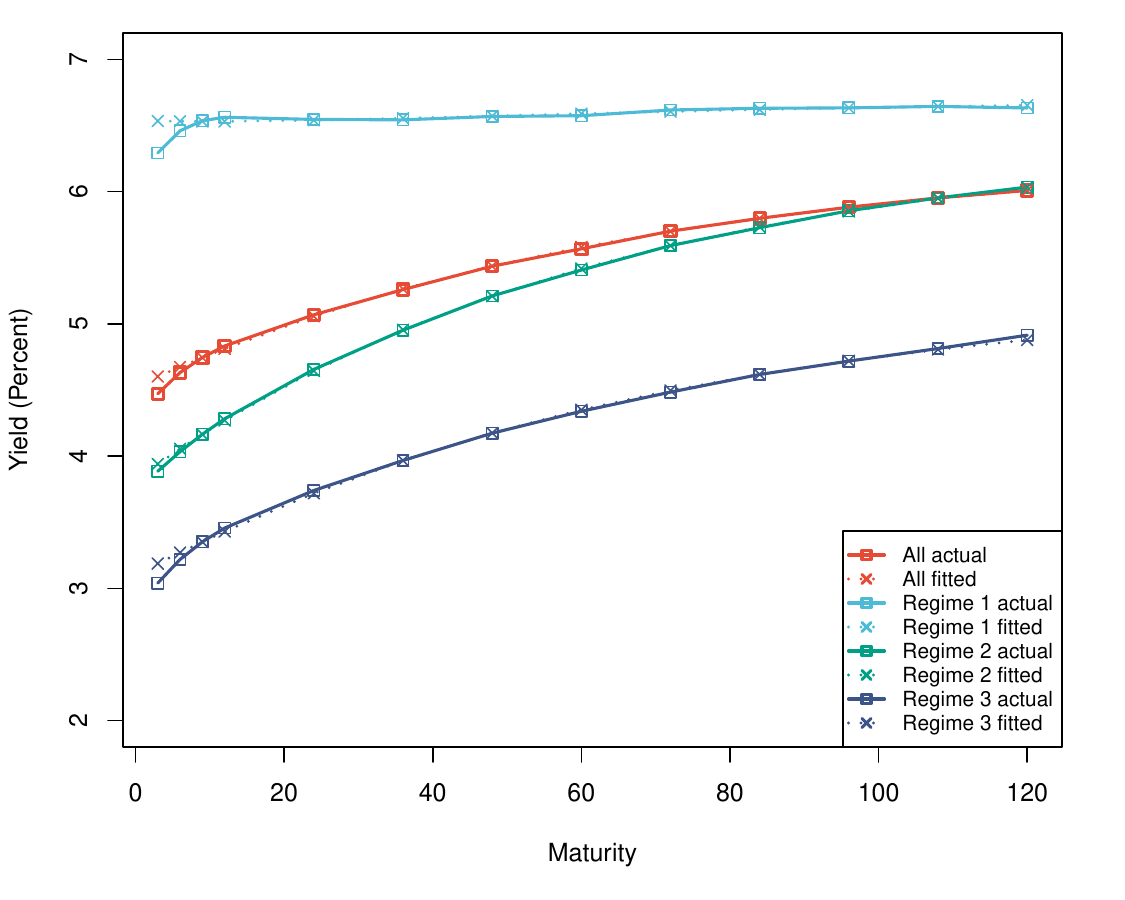}
		\captionsetup{skip=0pt}
		\label{fig:yield curve_yields-macro_3regime_case1}
		
	\end{center}
	
	\vspace{-4mm}
	
	\begin{spacing}{1.0}  \noindent  \footnotesize 
		Notes:  We show  actual yields  and fitted yield curves in each of three estimated regimes, as well as over the full sample (``all actual" and ``all fitted").
	\end{spacing}
	
	\bigskip
	
\end{figure}

In Figure \ref{fig:yield curve_yields-macro_3regime_case1} we compare aspects of the average actual and fitted yield curves. In red we show the pair of yield curves averaged across all regimes. The actual and fitted curves are very close, indeed almost identical at all but the shortest maturities, so the model fitting errors are generally very small. The remaining three pairs of curves correspond to the three regimes. First and very importantly, \textit{within} each regime, the actual and fitted curves are very close, again almost identical at all but the shortest maturities. Second and equally importantly, \textit{across} the regimes, the three  curve pairs are very different, underscoring the presence of three regimes and the distortions that would likely be caused if one were to insist on a one-regime model, as confirmed in the one-regime estimation results of Table \ref{tab:para_empirical_macroyields_1regime}, in which macro factors have no significant effects on yield factors.

\begin{table}[tb]
	\caption{Parameter Estimates, Single-Regime Yields-Macro Model}
	\label{tab:para_empirical_macroyields_1regime}
	
	\begin{center}
		
		\vspace{-5mm}
		
		\scriptsize
		\setlength{\tabcolsep}{1.1mm}{
			\begin{tabular*}{\textwidth}{@{\extracolsep{\fill}}lcccccc|cccccc|c}
				\toprule
				
				& \multicolumn{6}{c|}{$\A$}                                                                                                & \multicolumn{6}{c|}{$\HH$}                                                                             & $\boldsymbol{\mu}$ \\ \midrule
				& $L_{t-1}$ & $S_{t-1}$ & $C_{t-1}$ & $\variablename{CU}_{t-1}$ & $\variablename{FFR}_{t-1}$ & $\variablename{INFL}_{t-1}$ & $L_{t}$ & $S_{t}$ & $C_{t}$ & $\variablename{CU}_t$ & $\variablename{FFR}_t$ & $\variablename{INFL}_t$ &                    \\ \cmidrule{2-14} 
				\multirow{2}{*}{$L_{t}$}                 & 0.99*     & 0.00      & 0.01      & 0.00                      & 0.01                       & 0.00                        & 0.15*   & -0.05*  & -0.14*  & 0.01                  & 0.00*                  & 0.01*                   & 6.19*              \\
				& (0.01)    & (0.01)    & (0.01)    & (0.02)                    & (0.03)                     & (0.00)                      & (0.42)  & (0.37)  & (0.02)  & (0.02)                & (0.38)                 & (0.01)                  & (0.15)             \\
				\multirow{2}{*}{$S_{t}$}                 & 0.00      & 0.96*     & 0.01      & 0.01                      & 0.00                       & 0.00                        &         & 0.34*   & 0.05    & 0.05*                 & 0.09*                  & 0.01                    & -1.74*             \\
				& (0.01)    & (0.01)    & (0.01)    & (0.02)                    & (0.03)                     & (0.00)                      &         & (0.34)  & (0.03)  & (0.02)                & (0.34)                 & (0.01)                  & (0.12)             \\
				\multirow{2}{*}{$C_{t}$}                 & 0.00      & 0.00      & 0.92*     & 0.00                      & 0.01                       & 0.01                        &         &         & 1.03*   & 0.02                  & -0.10*                 & 0.00                    & -0.80              \\
				& (0.02)    & (0.02)    & (0.02)    & (0.00)                    & (0.02)                     & (0.02)                      &         &         & (0.07)  & (0.02)                & (0.03)                 & (0.01)                  & (0.43)             \\
				\multirow{2}{*}{$\variablename{CU}_t$}   & 0.29*     & 0.22*     & 0.02      & 0.99*                     & -0.26*                     & 0.00                        &         &         &         & 0.64*                 & 0.07*                  & 0.06*                   & 78.22*             \\
				& (0.08)    & (0.07)    & (0.02)    & (0.01)                    & (0.07)                     & (0.01)                      &         &         &         & (0.02)                & (0.03)                 & (0.01)                  & (0.08)             \\
				\multirow{2}{*}{$\variablename{FFR}_t$}  & 0.44*     & 0.44*     & 0.00      & 0.00                      & 0.60*                      & 0.00                        &         &         &         &                       & 0.23*                  & 0.02*                   & 4.78*              \\
				& (0.04)    & (0.04)    & (0.00)    & (0.02)                    & (0.03)                     & (0.00)                      &         &         &         &                       & (0.36)                 & (0.01)                  & (0.06)             \\
				\multirow{2}{*}{$\variablename{INFL}_t$} & 0.00      & 0.00      & -0.01     & 0.01*                     & 0.00                       & 0.99*                       &         &         &         &                       &                        & 0.08*                   & 3.73*              \\
				& (0.00)    & (0.00)    & (0.01)    & (0.00)                    & (0.00)                     & (0.00)                      &         &         &         &                       &                        & (0.01)                  & (0.04)             \\  \bottomrule
			\end{tabular*}
		}
		
	\end{center}
	
	\vspace{-4mm}

	\begin{spacing}{1.0}  \noindent  \footnotesize 
		
	Notes:  We show posterior means and standard deviations in each of the three regimes. Asterisks indicate that the posterior 95\% credible interval does not include 0. Because $\HH$ is a symmetric matrix, we show only its diagonal and upper-right elements.
	
\end{spacing}
	
	\bigskip
	
\end{table}

\subsection{Comparing Regimes}

Now we explore the three detected regimes in greater detail. We focus on factor dynamics, which are determined by the transition matrix $\A$ and factor innovation covariance matrix $\HH$. We first compare  posterior distributions of elements of the $\A$ and $\HH$ matrices across regimes (including formal tests for equal posterior means), and then we compare impulse-response functions.

\subsubsection{Posterior Distributions}

Here we compare posterior densities of estimated $\A$ and $\HH$ parameters across different regimes. Given their importance in determining the factor dynamics, we focus on the diagonal elements, denoted by $A_{i,i}$ and $H_{i,i}$, respectively.

\begin{figure}[tp]
	\caption{Posterior Densities of $\A$ and $\HH$,  Three-Regime Yields-Macro Model}    \label{postAH_density_yields_macro_3regimes_case1}
	
	\begin{center}
		
		%	\centering
		\begin{subfigure}{0.32\linewidth}
			\centering
			\includegraphics[width=0.9\linewidth]{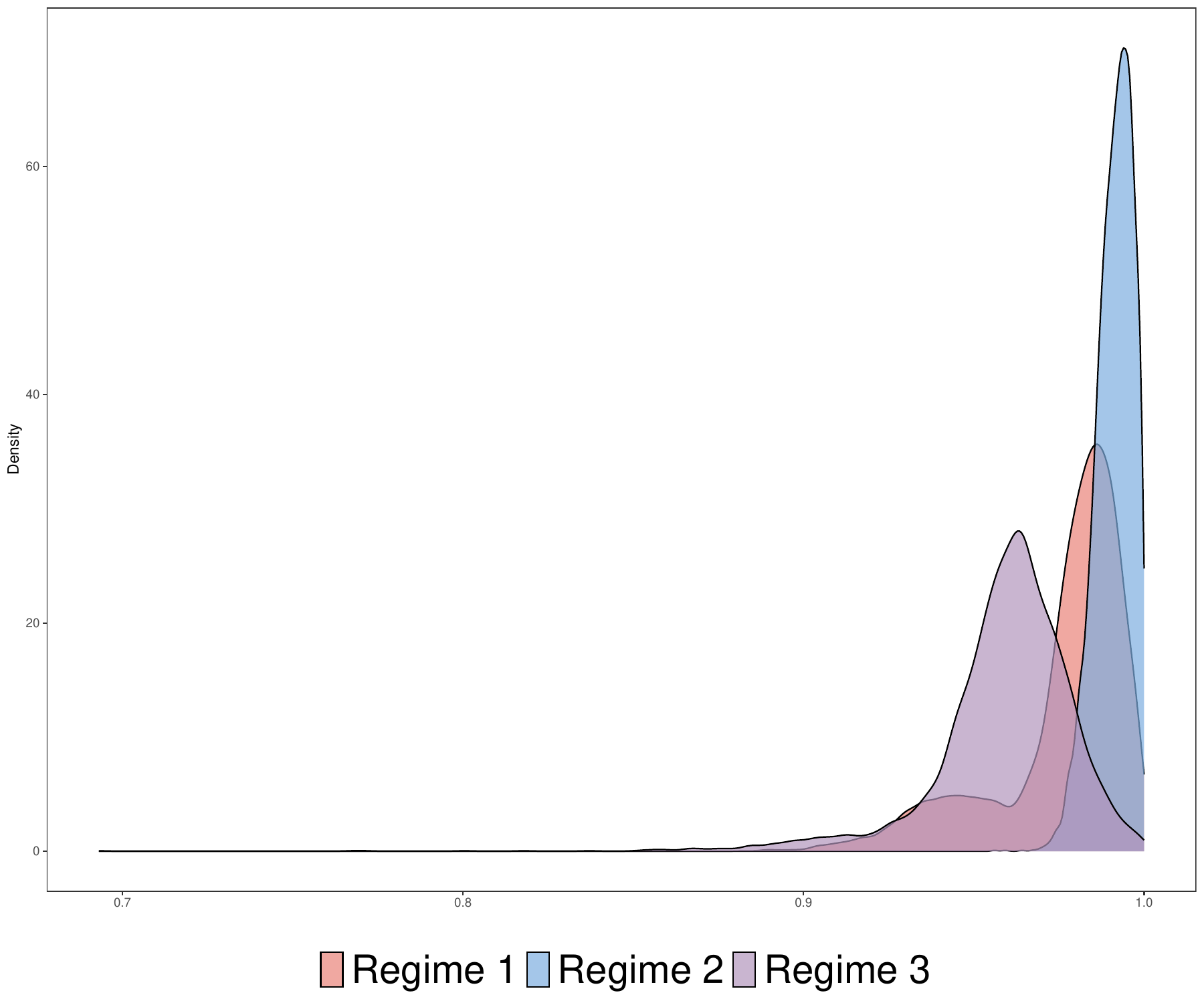}
			\caption{$A_{1,1}$}
			\label{A11_yields_macro_3regimes_case1}
		\end{subfigure}
		\centering
		\begin{subfigure}{0.32\linewidth}
			\centering
			\includegraphics[width=0.9\linewidth]{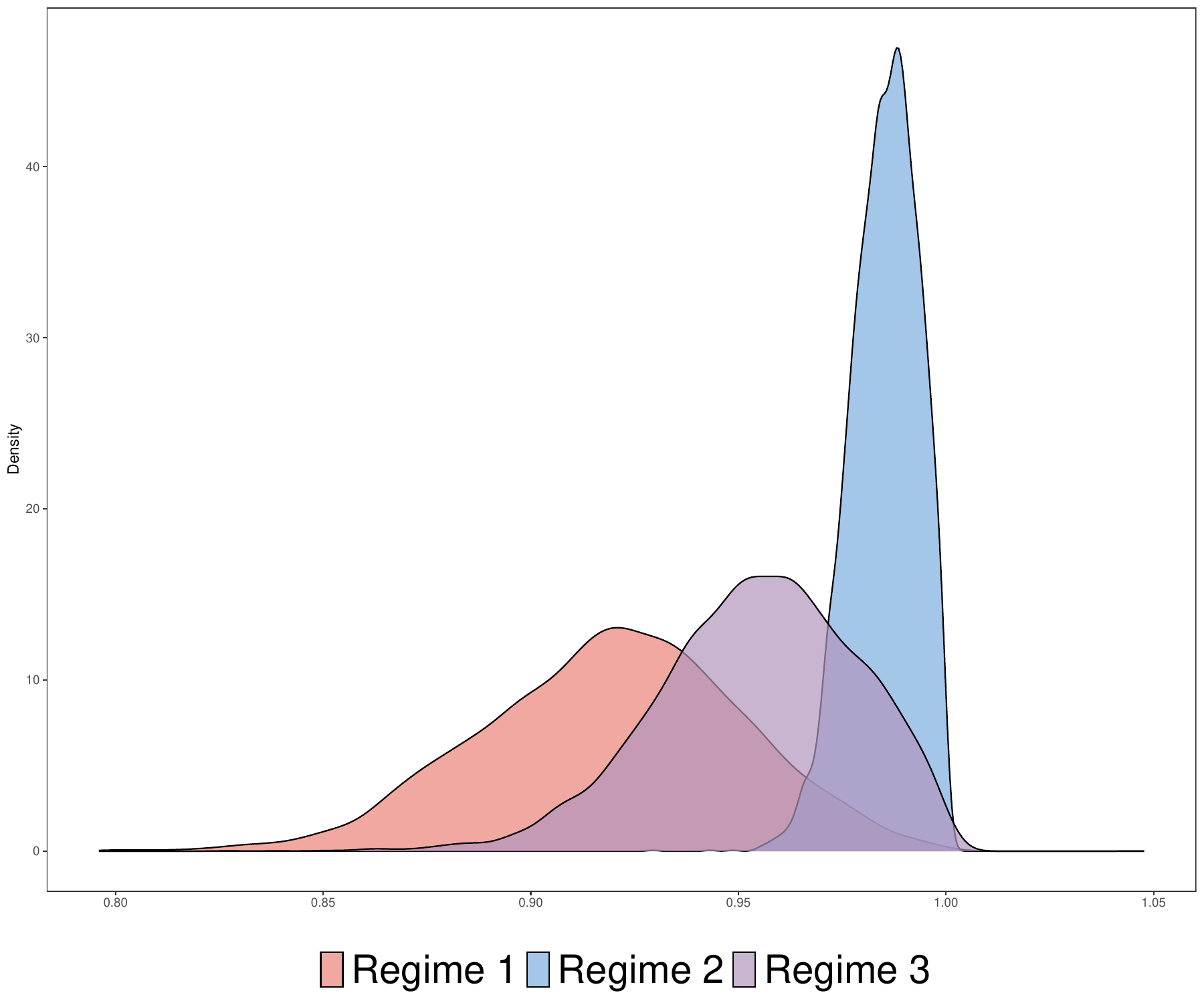}
			\caption{$A_{2,2}$}
			\label{A22_yields_macro_3regimes_case1}
		\end{subfigure}
		\centering
		\begin{subfigure}{0.32\linewidth}
			\centering
			\includegraphics[width=0.9\linewidth]{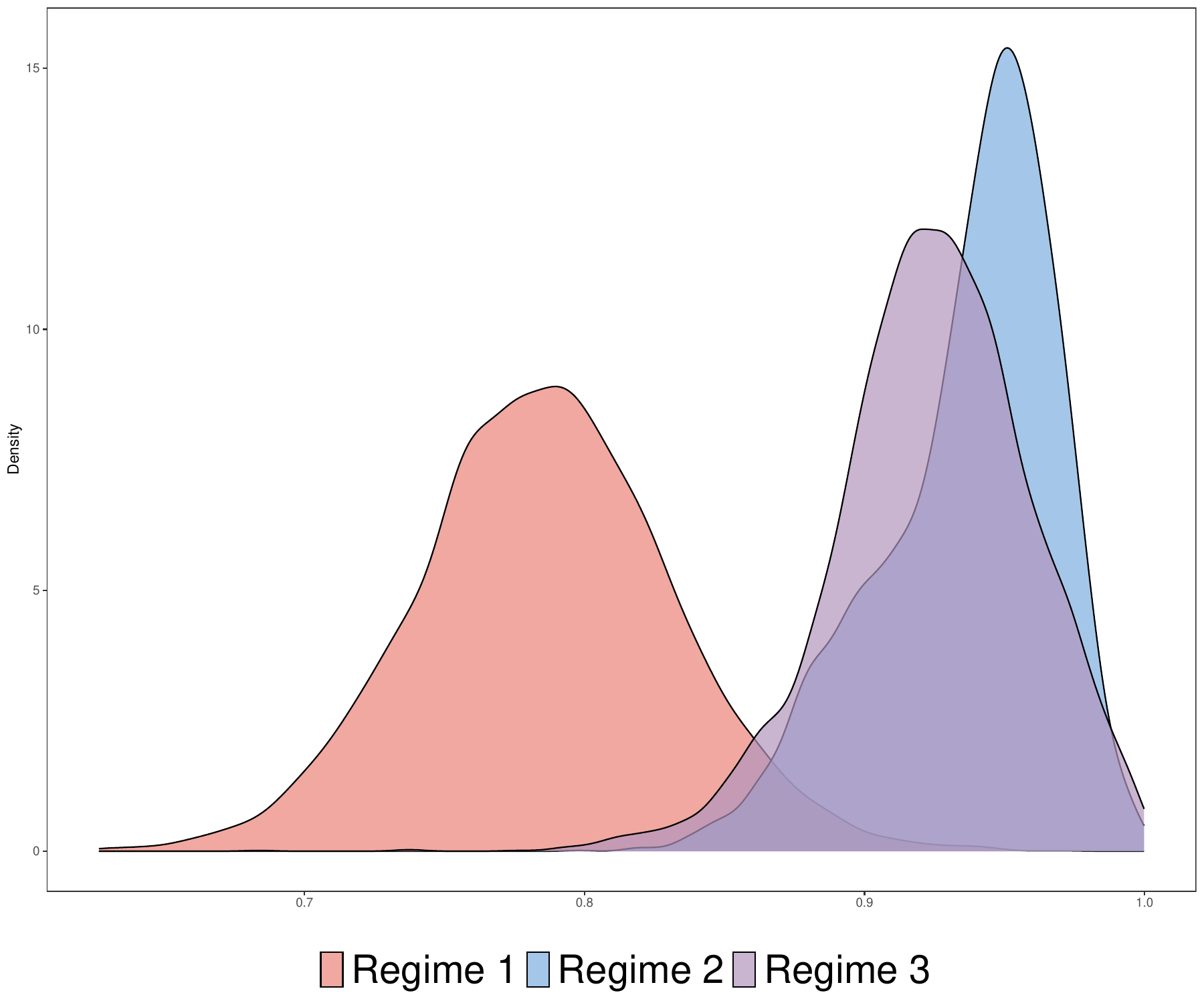}
			\caption{$A_{3,3}$}
			\label{A33_yields_macro_3regimes_case1}
		\end{subfigure}
		
		\centering
		\begin{subfigure}{0.32\linewidth}
			\centering
			\includegraphics[width=0.9\linewidth]{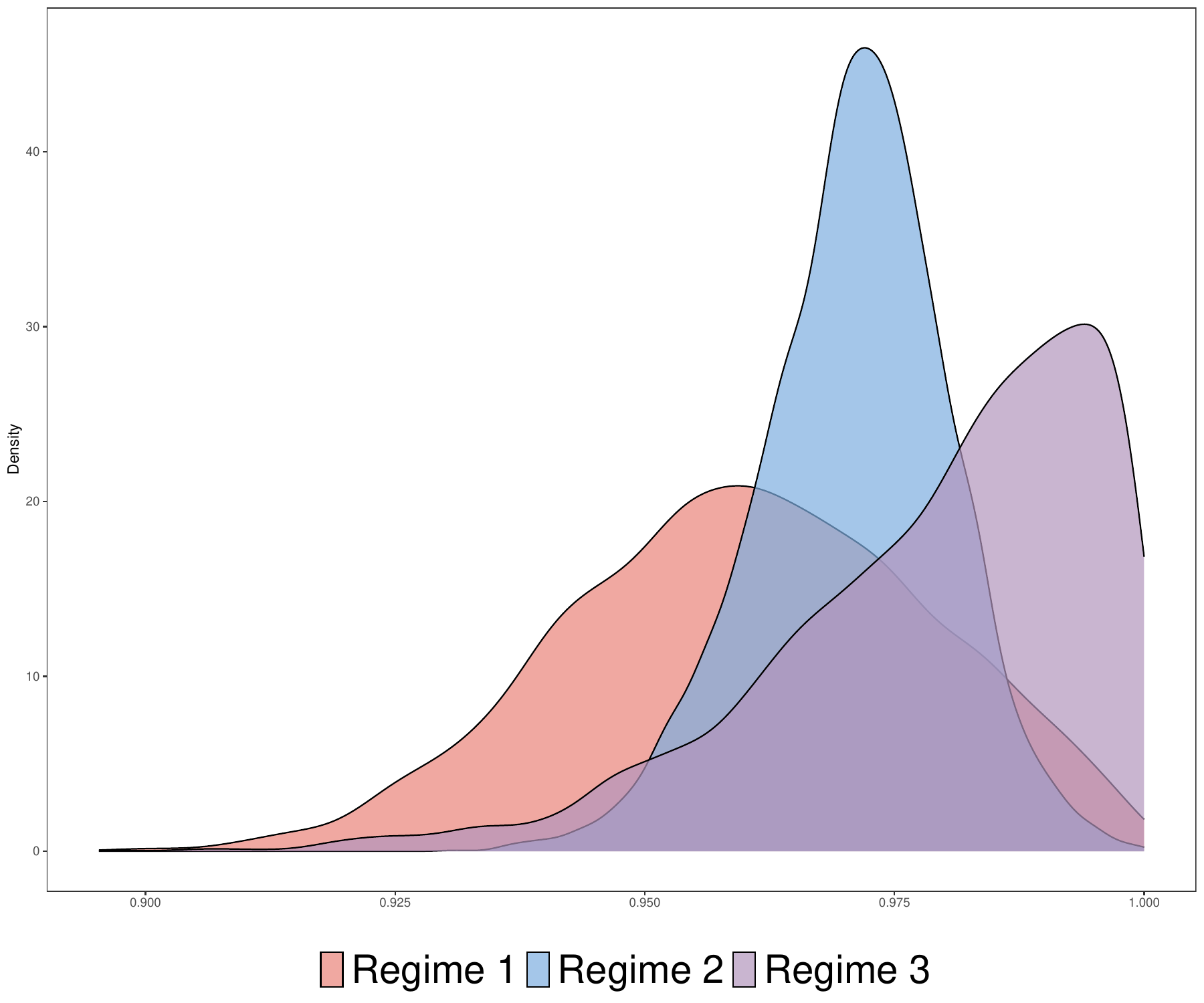}
			\caption{$A_{4,4}$}
			\label{A44_yields_macro_3regimes_case1}
		\end{subfigure}
		\centering
		\begin{subfigure}{0.32\linewidth}
			\centering
			\includegraphics[width=0.9\linewidth]{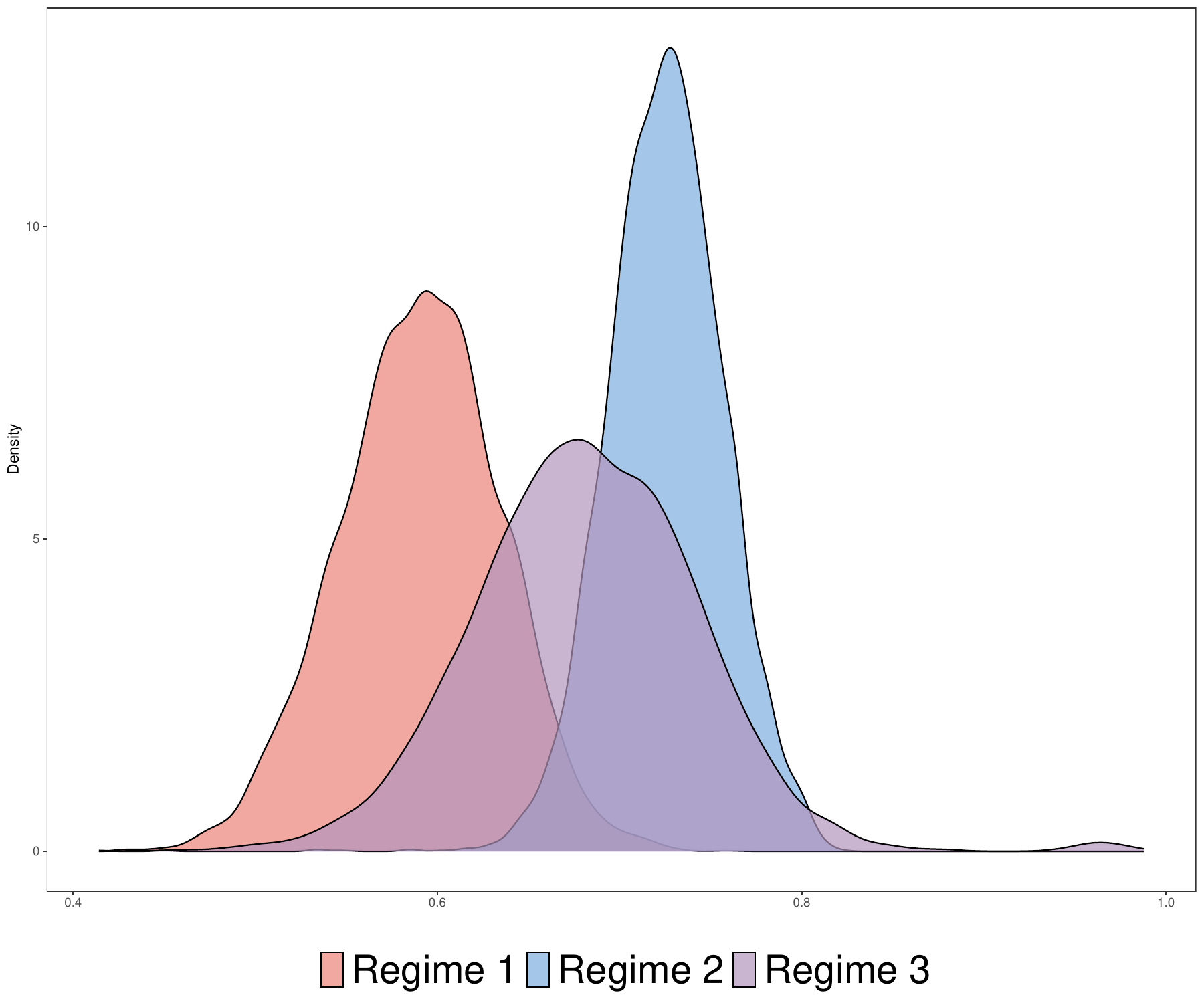}
			\caption{$A_{5,5}$}
			\label{A55_yields_macro_3regimes_case1}
		\end{subfigure}
		\centering
		\begin{subfigure}{0.32\linewidth}
			\centering
			\includegraphics[width=0.9\linewidth]{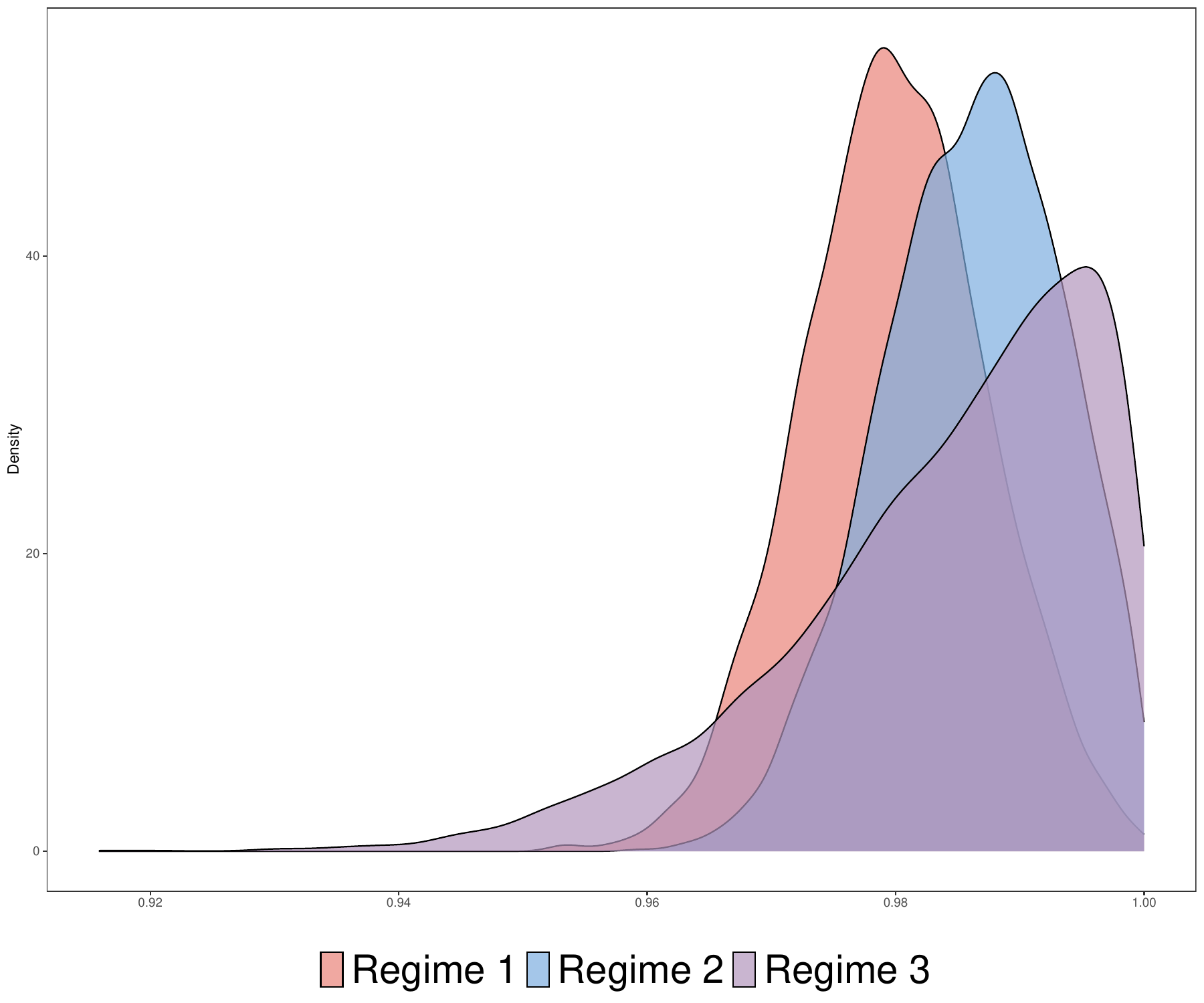}
			\caption{$A_{6,6}$}
			\label{A66_yields_macro_3regimes_case1}
		\end{subfigure}
		
		\centering
		\begin{subfigure}{0.32\linewidth}
			\centering
			\includegraphics[width=0.9\linewidth]{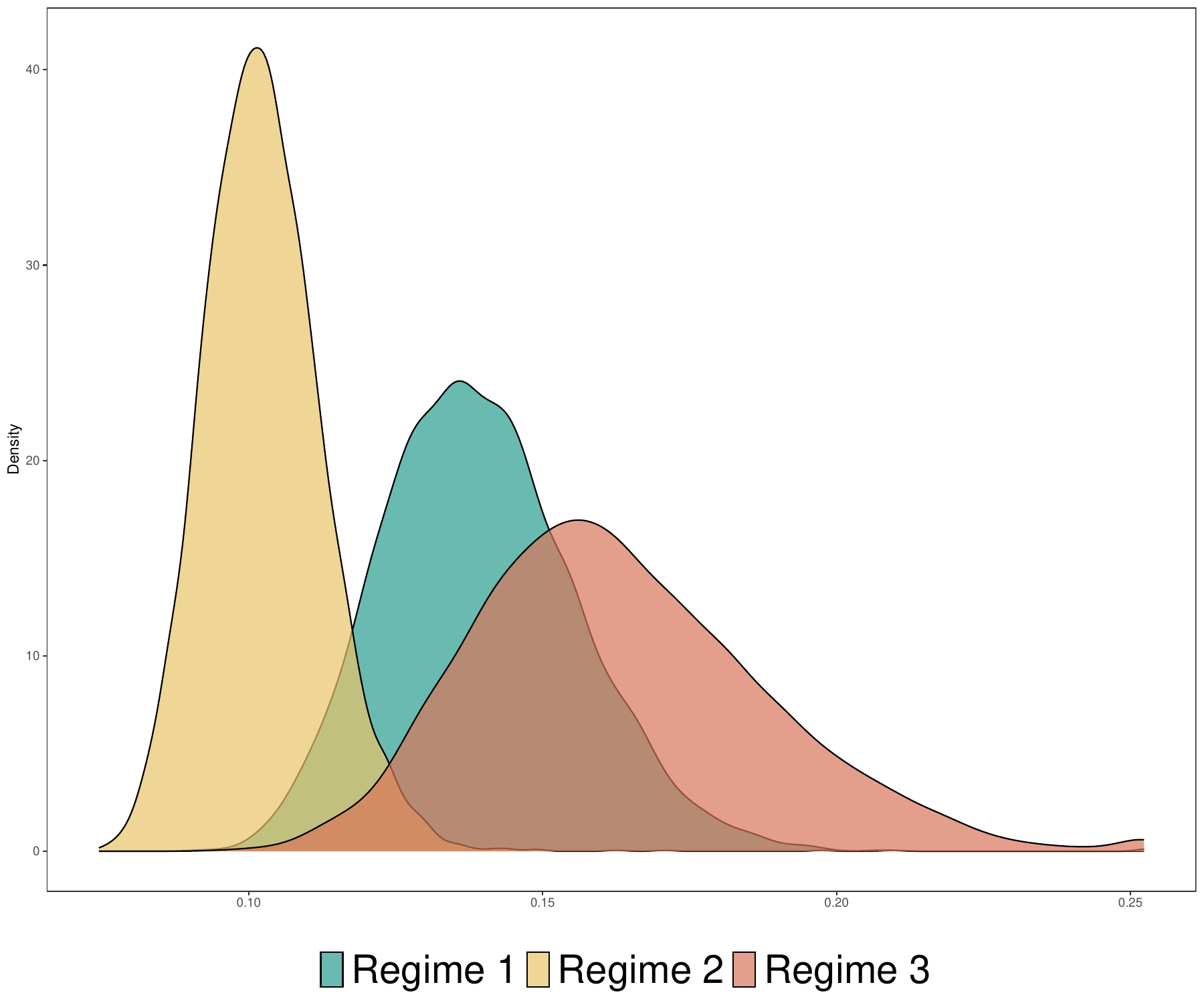}
			\caption{$H_{1,1}$}
			\label{H11_yields_macro_3regimes_case1}
		\end{subfigure}
		\centering
		\begin{subfigure}{0.32\linewidth}
			\centering
			\includegraphics[width=0.9\linewidth]{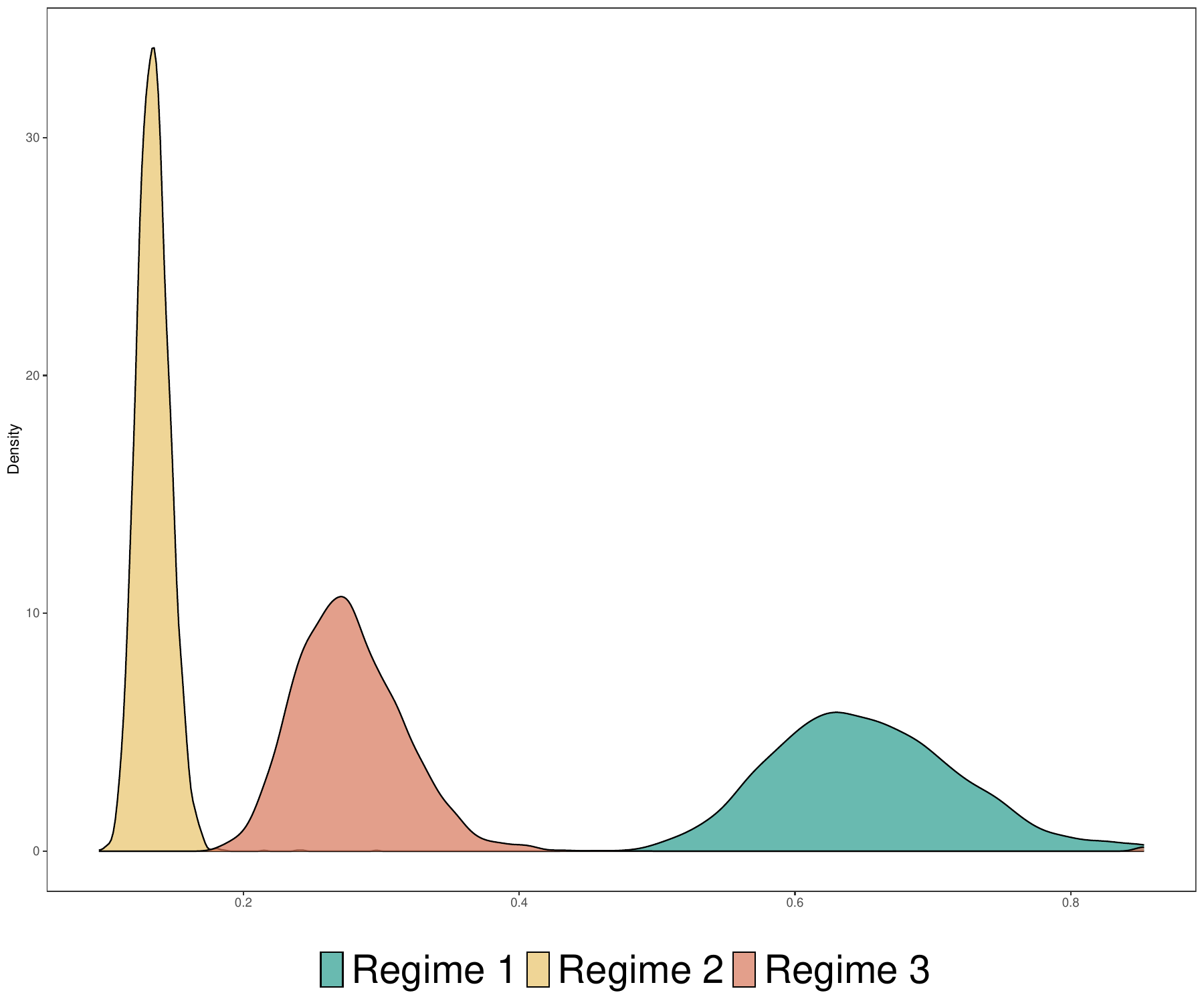}
			\caption{$H_{2,2}$}
			\label{H22_yields_macro_3regimes_case1}
		\end{subfigure}
		\centering
		\begin{subfigure}{0.32\linewidth}
			\centering
			\includegraphics[width=0.9\linewidth]{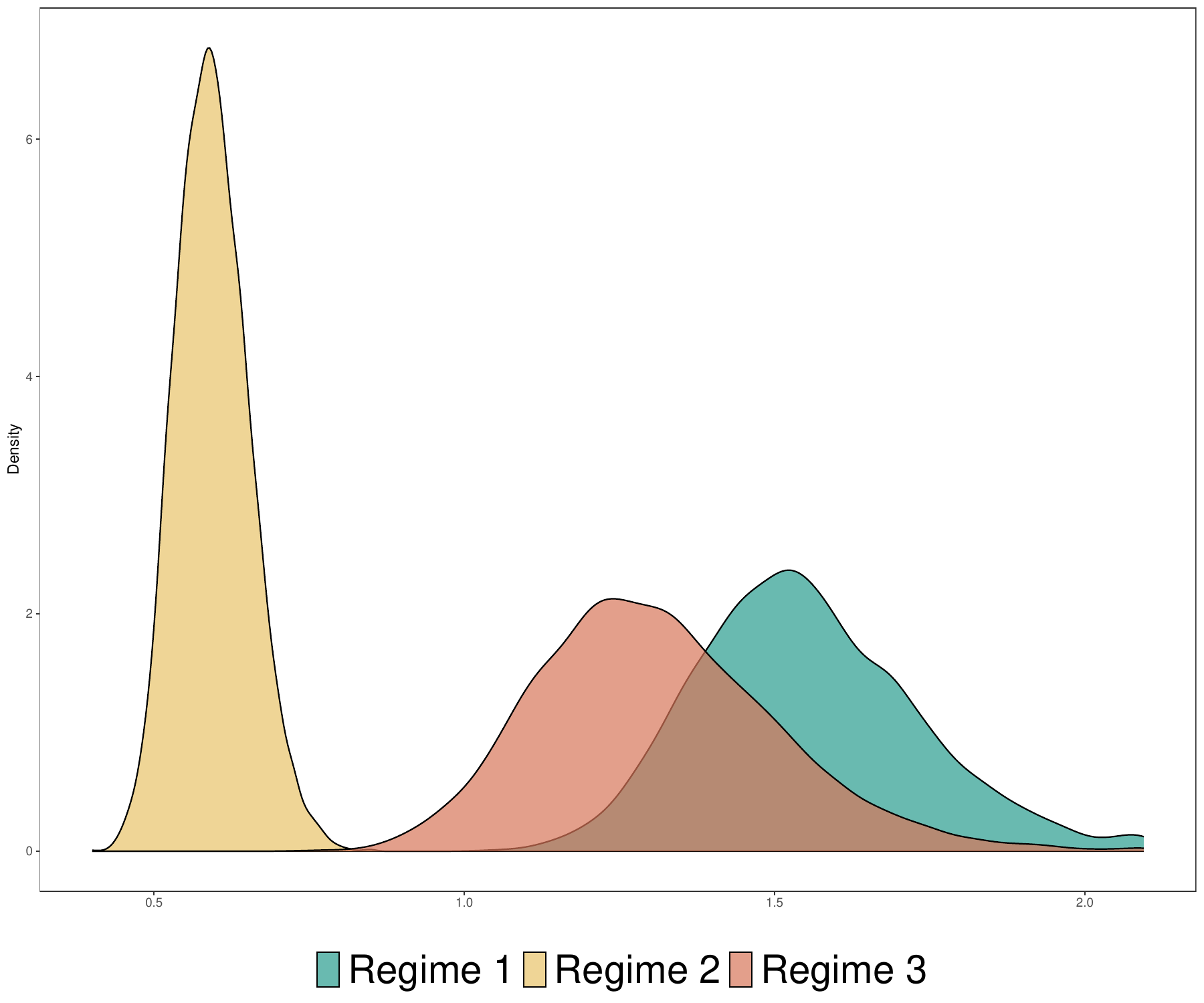}
			\caption{$H_{3,3}$}
			\label{H33_yields_macro_3regimes_case1}
		\end{subfigure}
		
		\centering
		\begin{subfigure}{0.32\linewidth}
			\centering
			\includegraphics[width=0.9\linewidth]{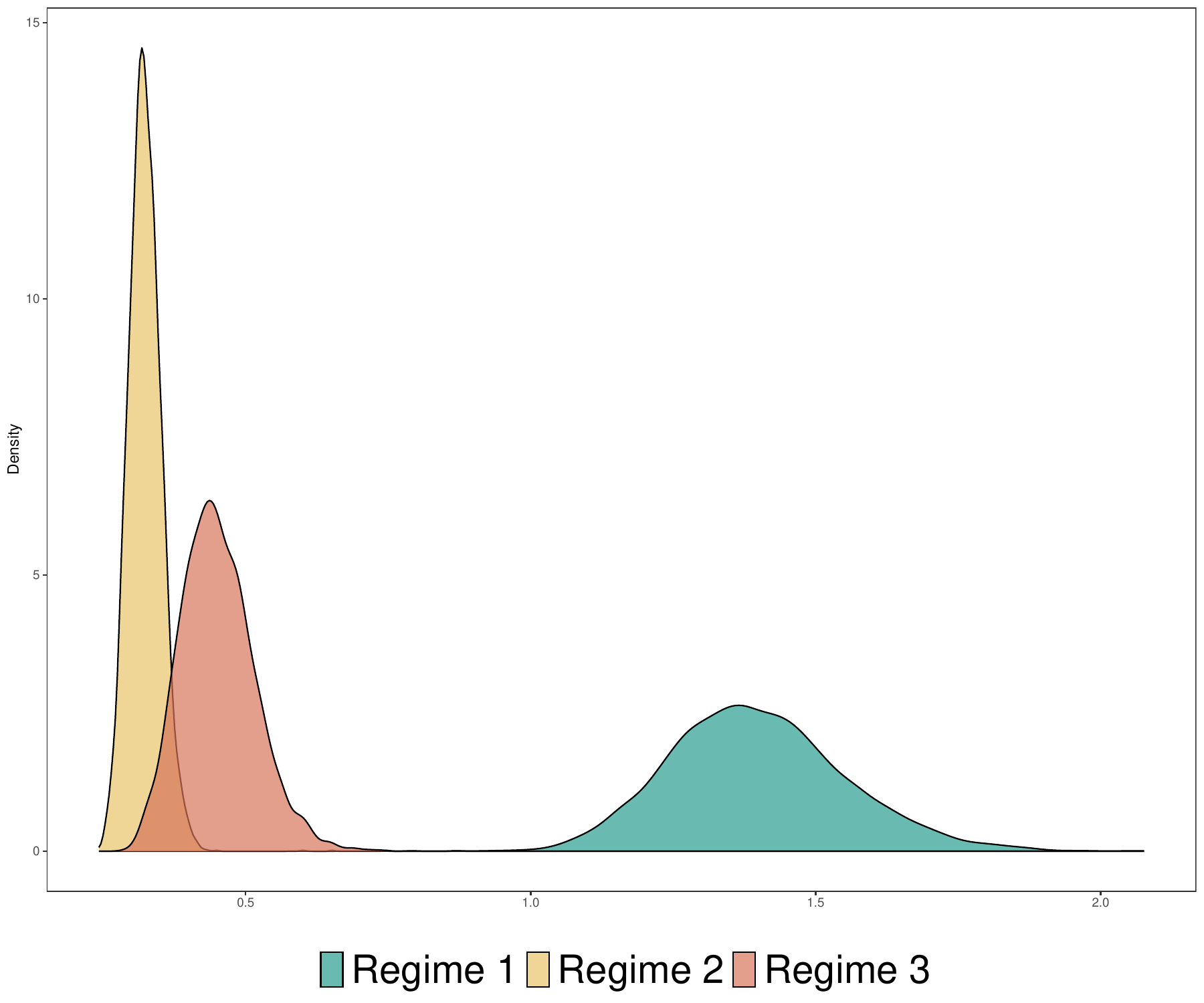}
			\caption{$H_{4,4}$}
			\label{H44_yields_macro_3regimes_case1}
		\end{subfigure}
		\centering
		\begin{subfigure}{0.32\linewidth}
			\centering
			\includegraphics[width=0.9\linewidth]{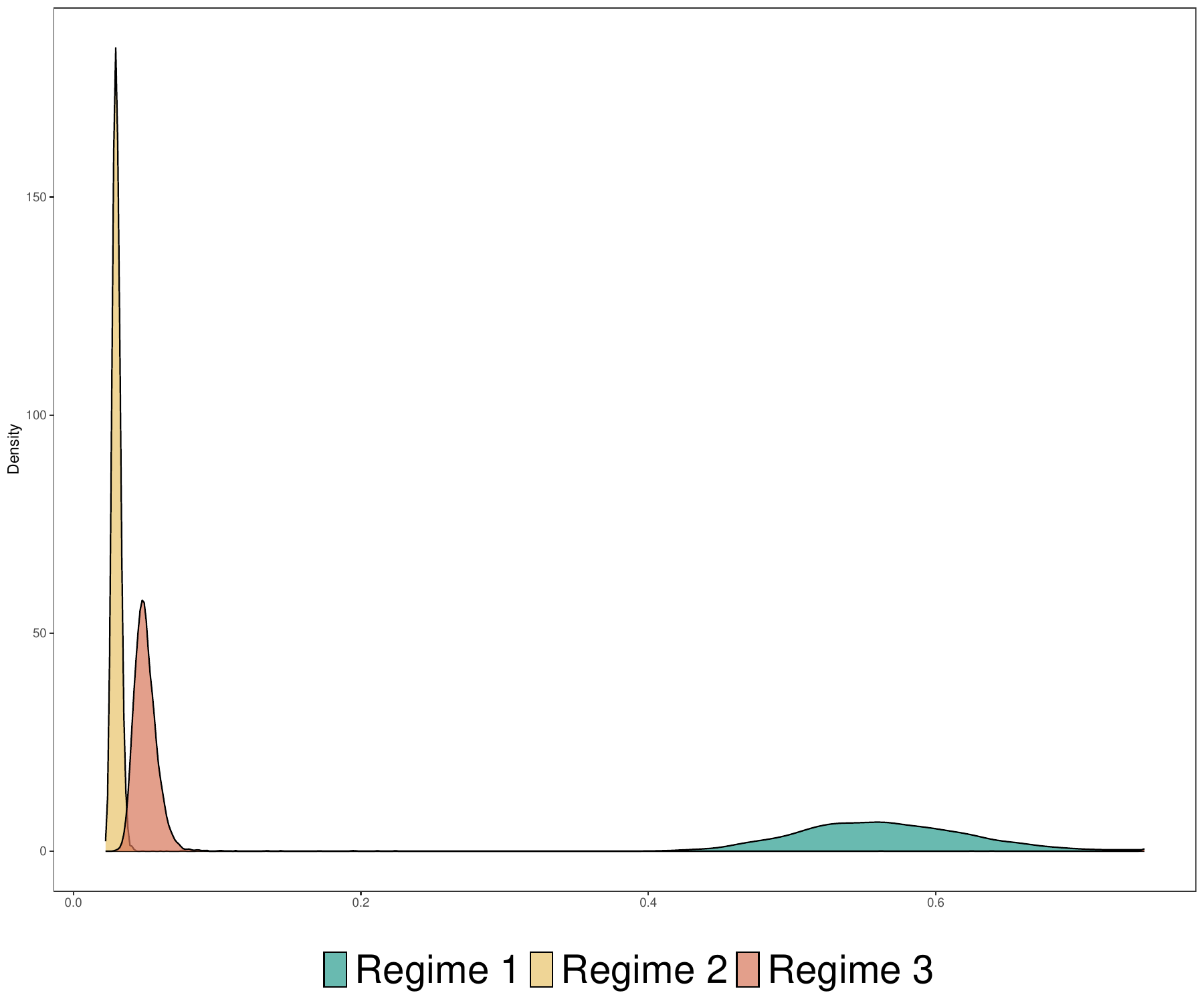}
			\caption{$H_{5,5}$}
			\label{H55_yields_macro_3regimes_case1}
		\end{subfigure}
		\centering
		\begin{subfigure}{0.32\linewidth}
			\centering
			\includegraphics[width=0.9\linewidth]{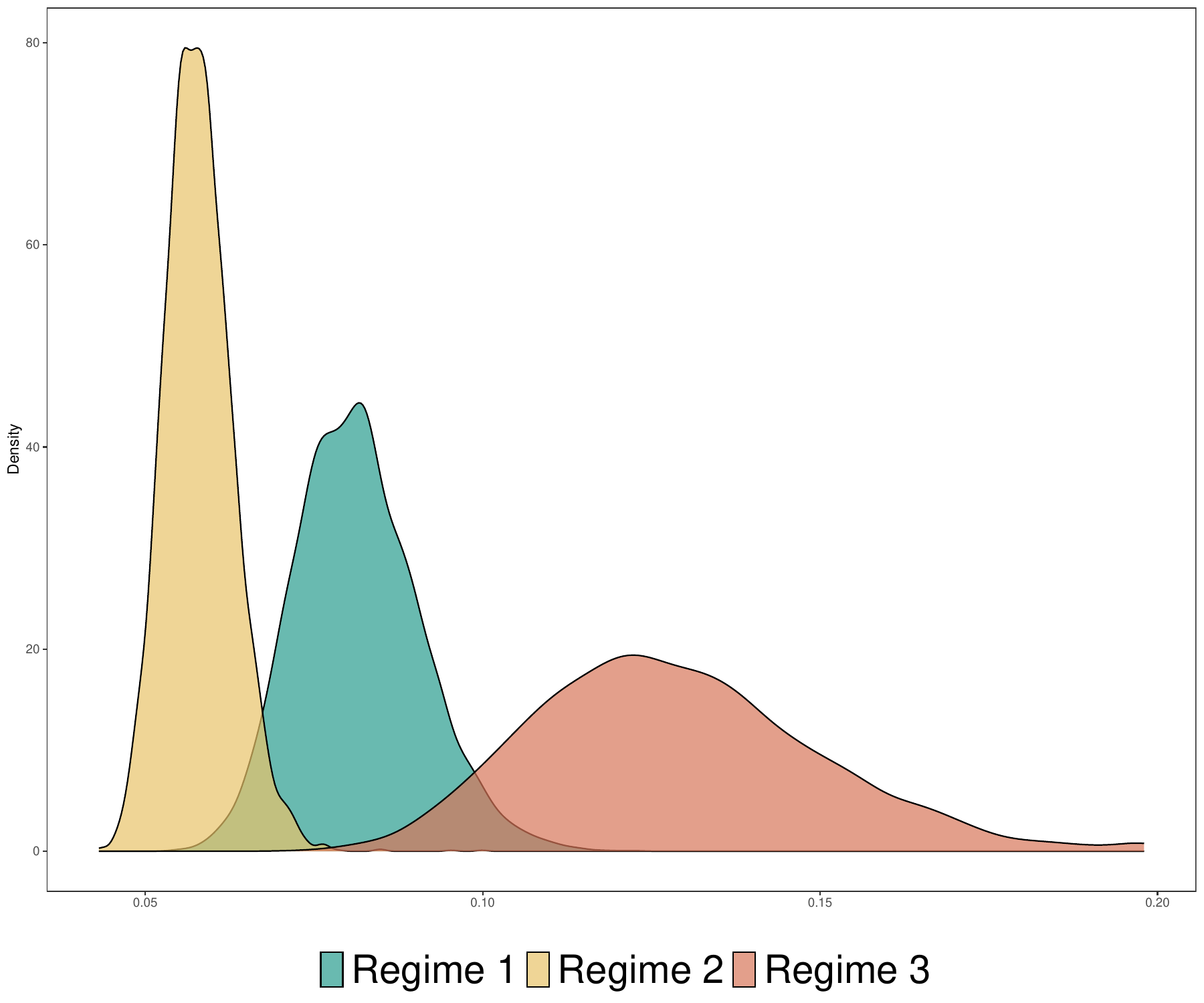}
			\caption{$H_{6,6}$}
			\label{H66_yields_macro_3regimes_case1}
		\end{subfigure}
		
	\end{center}
	
	\vspace{-3mm}
	
	\begin{spacing}{1.0}  \noindent  \footnotesize 
		Notes: We show posterior densities of the diagonal elements of the factor transition matrix $\A$ and covariance matrix $\HH$ for the yields-macro model with three regimes. Different colors indicate different regimes.
	\end{spacing}
	
\end{figure}

In Figure \ref{postAH_density_yields_macro_3regimes_case1} we show  posterior density plots for the yields-macro model parameters across the identified regimes. The plots reveal distinct posterior patterns for all parameters in each regime. For example, while Regimes 2 and 3 exhibit similar posterior densities for $A_{1,1}$ (representing level dynamics), they differ markedly for $A_{2,2}$ and $A_{3,3}$ (representing slope and curvature dynamics, respectively). Notably, the factor innovation covariance matrices also display substantial variation across regimes, with the posterior densities of the diagonal elements of $\HH$ showing minimal overlap, indicating important changes across regimes.

\begin{table}[tp]
	\caption{\textit{t}-tests, Three-Regime Yields-Macro Models}
	\label{tab:test_yields_macro}
	
	\vspace{-4mm}
	
	\begin{center}
		
		\footnotesize
		\begin{tabular*}{\textwidth}{@{\extracolsep{\fill}}cccccccc}
			\toprule
			\multirow{2}{*}{Model}         &  \multirow{2}{*}{Regimes}      & \multicolumn{3}{c}{$\A$}                                             & \multicolumn{3}{c}{$\HH$}                                                               \\ \cmidrule{3-5} \cmidrule{6-8}
			&                   &           & \textit{t}-statistic & \textit{p}-value &           & \textit{t}-statistic & \textit{p}-value \\ \midrule
			\multirow{18}{*}{Yields-Macro} & \multirow{3}{*}{1 vs 2}  & $A_{1,1}$ & -53.43                                & 0.00                              & $H_{1,1}$ & 132.67                                & 0.00                                 \\
			&                          & $A_{2,2}$ & -143.84                               & 0.00                              & $H_{2,2}$ & 530.22                                & 0.00                              \\
			&                          & $A_{3,3}$ & -193.58                               & 0.00                              & $H_{3,3}$ & 358.49                                & 0.00                              \\ \cmidrule{2-8} 
			& \multirow{3}{*}{1 vs 3}  & $A_{1,1}$ & 42.31                                 & 0.00                              & $H_{1,1}$ & -26.28                                & 0.00                              \\
			&                          & $A_{2,2}$ & -62.93                                & 0.00                              & $H_{2,2}$ & 233.38                                & 0.00                              \\
			&                          & $A_{3,3}$ & -170.93                               & 0.00                              & $H_{3,3}$ & 64.16                                 & 0.00                              \\ \cmidrule{2-8} 
			& \multirow{3}{*}{2 vs 3}  & $A_{1,1}$ & 107.34                                & 0.00                              & $H_{1,1}$ & -63.18                                & 0.00                              \\
			&                          & $A_{2,2}$ & 85.51                                 & 0.00                              & $H_{2,2}$ & -114.46                               & 0.00                              \\
			&                          & $A_{3,3}$ & 19.76                                 & 0.00                              & $H_{3,3}$ & -247.82                               & 0.00                              \\ \cmidrule{2-8} 
			& \multirow{3}{*}{1 vs 2}  & $A_{4,4}$ & -36.09                                & 0.00                              & $H_{4,4}$ & 484.88                                & 0.00                              \\
			&                          & $A_{5,5}$ & -176.16                               & 0.00                              & $H_{5,5}$ & 628.39                                & 0.00                              \\
			&                          & $A_{6,6}$ & -42.59                                & 0.00                              & $H_{6,6}$ & 159.72                                & 0.00                              \\ \cmidrule{2-8} 
			& \multirow{3}{*}{1 vs 3}  & $A_{4,4}$ & -54.90                                & 0.00                              & $H_{4,4}$ & 400.66                                & 0.00                              \\
			&                          & $A_{5,5}$ & -85.33                                & 0.00                              & $H_{5,5}$ & 398.66                                & 0.00                              \\
			&                          & $A_{6,6}$ & -20.26                                & 0.00                              & $H_{6,6}$ & -137.57                               & 0.00                              \\ \cmidrule{2-8} 
			& \multirow{3}{*}{2 vs 3}  & $A_{4,4}$ & -31.88                                & 0.00                              & $H_{4,4}$ & -128.02                               & 0.00                              \\
			&                          & $A_{5,5}$ & 43.61                                 & 0.00                              & $H_{5,5}$ & -24.06                                & 0.00                              \\
			&                          & $A_{6,6}$ & 10.07                                 & 0.00                              & $H_{6,6}$ & -216.70                               & 0.00                              \\ \bottomrule
		\end{tabular*}
		
	\end{center}
	
	\footnotesize
	
	Notes: We show \textit{t}-tests for the number of regimes in the yields-macro models.
	
	\bigskip
	
\end{table}

To further examine the differences among detected regimes, we implement two-sample \textit{t}-tests if inter-regime equality of posterior means of the elements of $\A$ and $\HH$, again focusing on their diagonal elements. The sample sizes are equal across regimes, reflecting the uniform posterior sample size $n_s$ used in the Gibbs sampler, but, given the potential for heterogeneous variances among regimes, we implement tests that accommodates unequal variances $s_1^2$ and $s_2^2$. The \textit{t}-statistic is
\begin{equation*}
		t = \frac{\bar{X_1}-\bar{X_2}}{s_{\Delta}},
\end{equation*}
where
$$s_{\Delta} = \sqrt{\frac{s_1^2 + s_2^2}{n_s}},$$ 
with associated degrees of freedom $\frac{(n_s-1)(s_1^2 + s_2^2)^2}{s_1^4 + s_2^4}$.  We present the \textit{t}-test results for the three-regime yields-macro DNS model in Table \ref{tab:test_yields_macro}. They clearly indicate that the three identified regimes are distinct.

\subsubsection{Impulse Response Functions}
 
\begin{figure}[p]
	\caption{Impulse Responses, Three-Regime Yields-Macro Model}
	\label{fig:irf_yields-macro_3regime_case1_regime1}
\begin{center}
\vspace{-2mm}
Regime 1\\
 \vspace{-3mm}
	 Yields to Macro \quad \quad  \quad \quad  \quad \quad  \quad \quad  \quad \quad  \quad \quad Macro to Yields\\~\\
\vspace{-3mm}
		\includegraphics[width=0.45\linewidth]{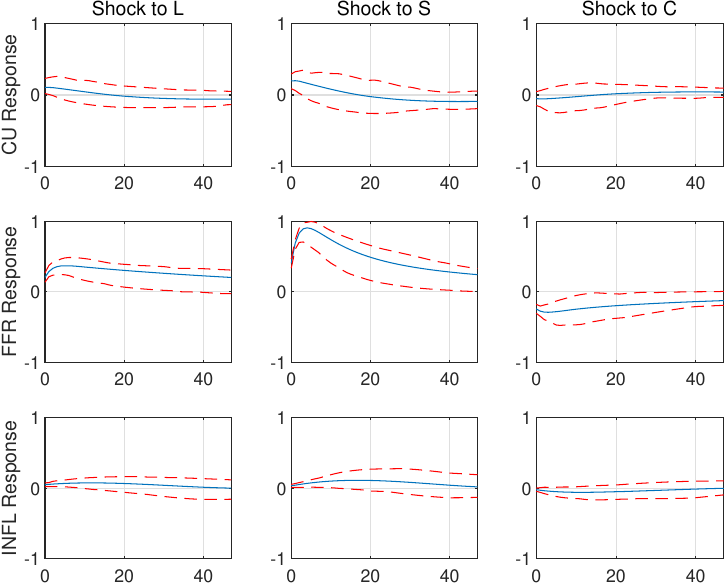} \quad \quad \quad
\includegraphics[width=0.45\linewidth]{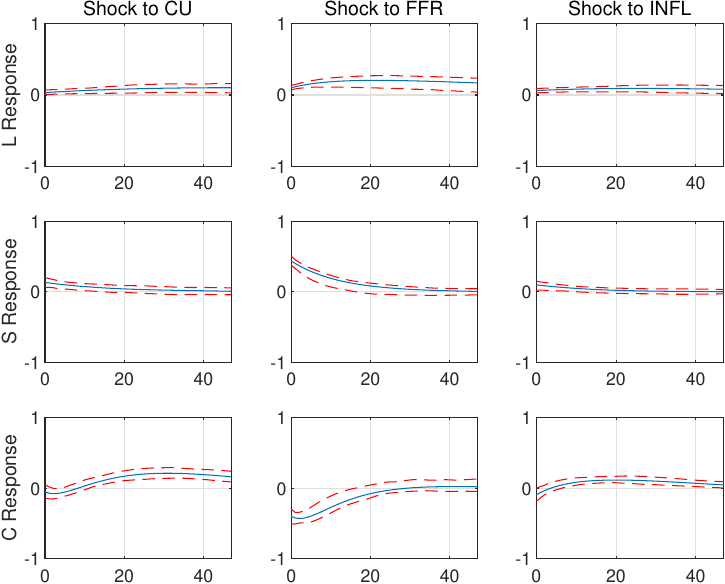}\\
		\bigskip
		
%		\bigskip
\vspace{-2mm}
Regime 2 \\
 \vspace{-3mm}
 Yields to Macro \quad \quad  \quad \quad  \quad \quad  \quad \quad  \quad \quad  \quad \quad Macro to Yields\\~\\
		\vspace{-3mm}
			\includegraphics[width=0.45\linewidth]{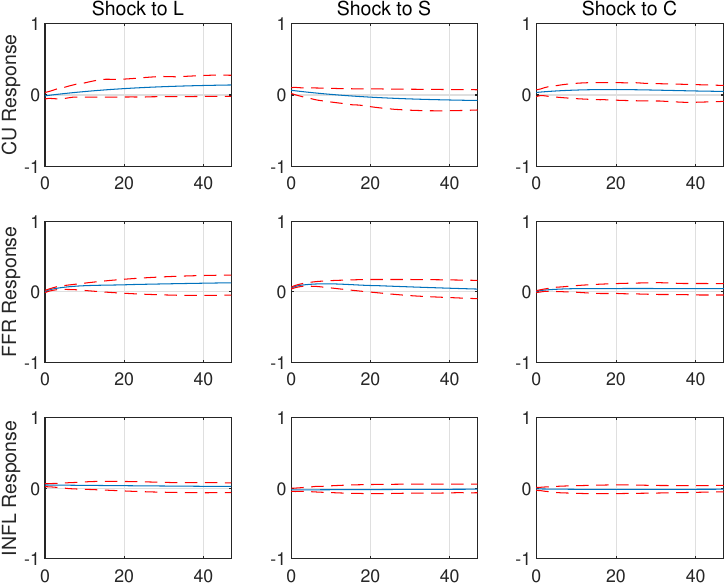}\quad \quad \quad 
			\includegraphics[width=0.45\linewidth]{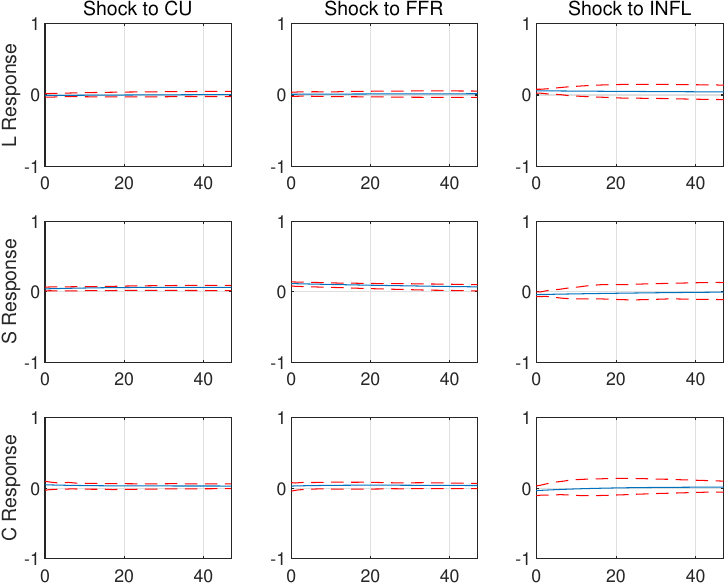} \\
%			\bigskip
					\bigskip
\vspace{-2mm}
Regime 3 \\
 \vspace{-3mm}
		Yields to Macro \quad \quad  \quad \quad  \quad \quad  \quad \quad  \quad \quad  \quad \quad Macro to Yields\\~\\
		\vspace{-3mm}
			\includegraphics[width=0.45\linewidth]{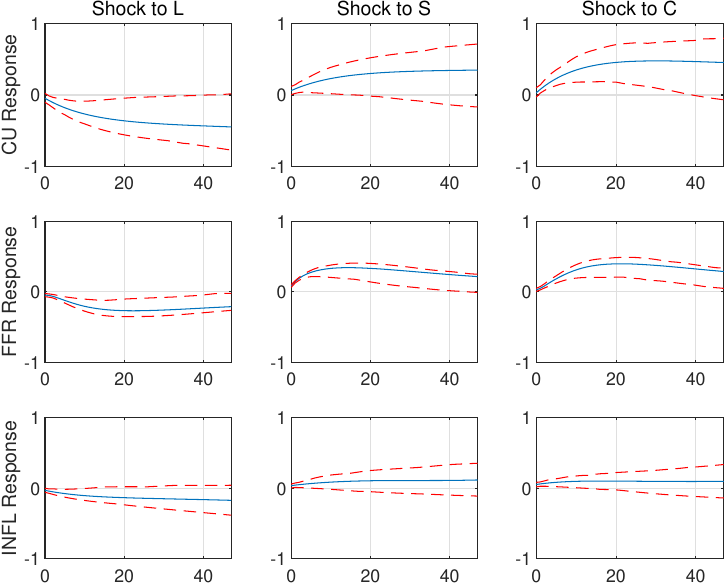} \quad \quad \quad 
	\includegraphics[width=0.45\linewidth]{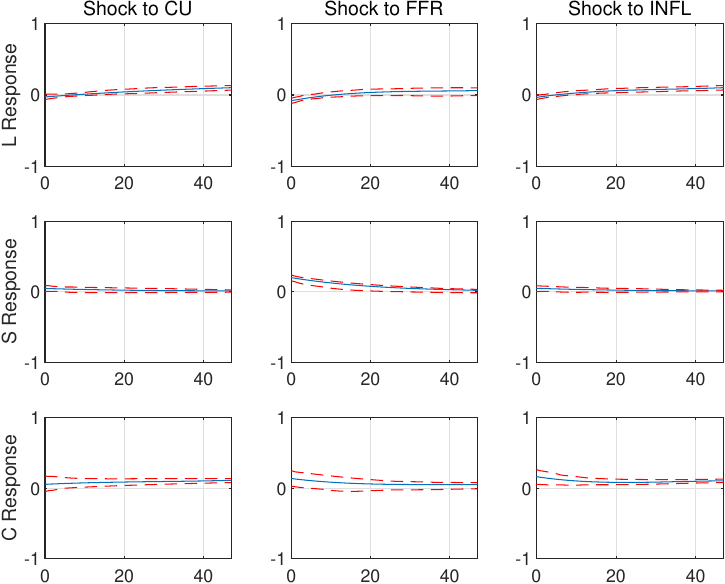}
	\end{center}
	\begin{spacing}{1.0}  \noindent  \footnotesize 
		Notes:  Point estimates are blue and 95\% credible bands are red.
	\end{spacing}

\bigskip
	
\end{figure}

Impulse response functions (IRFs) have been used at least since \cite{sims1980macroeconomics} to quantify the dynamic effects of shocks.  Our focus is on cross-regime differences in impulse responses in the three-regimes yields-macro DNS model, with particular attention to macro-spanning. In Figure  \ref{fig:irf_yields-macro_3regime_case1_regime1}, we show  ``yields-to-macro" and ``macro-to-yields" IRFs and credible bands.\footnote{We use generalized impulse response functions \citep{koop1996impulse} because they are invariant to variable ordering.}

Let us discuss the IRFs in reverse order of regimes, beginning with regime 3 (low $FFR$, high $INFL$).  In regime 3 the IRFs behave as expected under macro spanning.  That is, there are  clear yields-to-macro responses but little or no macro-to-yields responses. In regime 2 (low $FFR$, low $INFL$), the IRFs also support macro spanning insofar as there are no important macro-to-yields responses. At the same time, however, there are also no important yields-to-macro responses; that is, causal shock transmission is weak in both directions. Finally, in regime 1  (high $FFR$).  There is clear violation of macro spanning due to sizable macro-to-yields effects, with $FFR$ affecting all yield factors, and $CU$ affecting the curvature factor, again suggesting that macro spanning (or lack thereof) is a regime-specific phenomenon.

\section{Conclusion} \label{sec:conclusion}

We explore tree-based macroeconomic regime-switching in the context of the ``yields-macro" version of the dynamic Nelson-Siegel (DNS) yield-curve model. By integrating decision trees from machine learning, our approach customizes the tree-growing algorithm to partition macroeconomic variables based on the DNS model's Bayesian marginal likelihood, enabling us to identify regime-shifting patterns in the yield curve. Compared to traditional Markov-switching models, this ``macro instrumented" approach offers clear economic interpretation via macroeconomic linkages and ensures computational simplicity.

We provide a detailed empirical analysis of the U.S. Treasury yield curve, August 1971 - December 2022.  There are several findings. First, the macro-instrumented approach clearly identifies regime-switching behavior based on the yields-macro models, fitting the data well.  Second, the regimes are primarily driven by the federal funds rate, which captures the central bank (in our application, the U.S. Federal Reserve) policy stance. Third, we identify a regime with high federal funds rate, in which the  macroeconomy contains useful information about the future yield curve, whereas in other regimes the macroeconomy contains little or no information about the future yield curve.

Overall, then, our results indicate not only that yields clearly and consistently help to forecast the aspects of the macroeconomy, but also that aspects of the macroeconomy sometimes help to forecast yields. That is, our results suggest that ``macro spanning" may hold in some regimes but fail in others. Financial economic and econometric yield-curve modeling may therefore benefit from a more nuanced  re-orientation, working not in environments where macro spanning ``holds" or ``fails", but rather in environments where it sometimes holds and sometimes fails.

\appendix
\appendixpage
 
\newcounter{saveeqn}
\setcounter{saveeqn}{\value{section}}
\renewcommand{\theequation}{\mbox{\Alph{saveeqn}\arabic{equation}}} \setcounter{saveeqn}{1}
\setcounter{equation}{0}

\section{Posterior Inference}
\label{sec:gibbs}

\renewcommand{\thefigure}{A\arabic{figure}}
\setcounter{figure}{0}

\renewcommand{\thetable}{A\arabic{table}}
\setcounter{table}{0}

This section introduces the details of the Gibbs sampler for posterior inference. First we clarify notations: let $\tilde{\y}_T = [\y_1, \cdots, \y_T]^T$, $\tilde{\F}_T = [\F_1,\cdots, \F_T]^T$ be the collection of all data observations and latent factors. Let $\boldsymbol{\Omega} = [\A_0, \A_1, \A_2, \HH_0, \HH_1, \HH_2]$ represent all parameters of the regression coefficients and residual covariance matrices. Suppose the yield data is balanced, $N$ bonds for $T$ months and $n = N\times T$ total observations. Furthermore consider $K$ macro factors in addition to the $3$ factors of dynamic Nelson-Siegel model, $K$ can be zero for the case without macroeconomic variable. The dimension of all matrices are listed as follows $\y_t, \boldsymbol{\varepsilon}_t$: $N\times 1$, $\boldsymbol{\Lambda}$: $N\times 3$, $\f_t, \F_t, \boldsymbol{\mu}_0, \boldsymbol{\mu}_1, \boldsymbol{\mu}_2, \boldsymbol{\eta}_t$: $(3+K)\times 1$, $\A_0, \A_1, \A_2$: $(3+K)\times (3+K)$, $\Q$: $N\times N$, $\HH_0, \HH_1, \HH_2$: $(3+K)\times (3+K)$. The initial values for parameters are obtained from two step approach results. Initial values of Kalman filter $\F_0$ and $\mathbf{P}_0$ are equal to estimates by using Econometrics Toolbox state-space models (SSM) in MATLAB.
The prior for parameters and full Gibbs sampler for posterior inference are as follows.
\begin{itemize}
    \item Prior specification for different parameters
    \begin{itemize}
        \item Spike-and-slab prior: $\xi_0^2 = 10^{-5}$, $\xi_1^2 = 1$.
        \item Decay parameter $\lambda$ prior: $a = 0.01$, $b = 0.1$.
        \item Diagonal elements $\sigma^2_i$ of $\Q$ prior: $\alpha = 5$, $\beta = 0.05$.
        \item Factor mean $\bmu$ prior: $\underline{\boldsymbol{\mu}}$ is equal to initial value of $\bmu$, $\underline{\B} = \text{diag}(10)_{3+K}$.
        \item Covariance matrix $\HH$ prior: $\m_0$ and $\M_0$ are set according to initial value of $\HH$, i.e., prior mean closing to initial values. 
    \end{itemize}
\end{itemize}

\begin{itemize}
	\item $\tilde{\F}_T \mid \boldsymbol{\Omega}, \tilde{\y}_T, \boldsymbol{\Lambda}, \Q, \boldsymbol{\mu}_0, \boldsymbol{\mu}_1, \boldsymbol{\mu}_2$.
	      Sampling latent factors $\tilde{\F}_t$ conditional on all other parameters is achieved by the Kalman filter/smoother. Given $\boldsymbol{\Omega}$ and $\F_0$, the posterior distribution of $\tilde{\F}_T$ can be decomposed as a set of conditionals
	      \begin{equation*}
		      p(\tilde{\F}_T \mid \tilde{\y}_T) = p(\F_T \mid \tilde{\y}_T) \prod_{t=1}^{T-1}p(\F_t \mid \F_{t+1}, \tilde{\y}_t).
	      \end{equation*}
	      The above equation suggests that the factors $\tilde{\F}_T$ can be sampled sequentially by the Kalman filter, where all the conditionals are defined as
	      \begin{equation*}
		      \begin{aligned}
			      \widehat{\F}_{t\mid t-1} & =\A_{z_{t-1}}\F_{t-1 \mid t-1}                                                                 \\
			      \widehat{\mathbf{P}}_{t\mid t-1} & =\A_{z_{t-1}} \mathbf{P}_{t-1\mid t-1} \A_{z_{t-1}}^{T}+\HH_{z_t}                                         \\
			      \K_t                     & =\widehat{\mathbf{P}}_{t \mid  t - 1} \boldsymbol{\Lambda}^{T}(\boldsymbol{\Lambda} \widehat{\mathbf{P}}_{t\mid t-1} \boldsymbol{\Lambda}^{T}+\Q)^{-1} \\
			      \F_{t\mid t}             & =\widehat{\F}_{t \mid t-1}+\K_t(\y_{t}-\boldsymbol{\Lambda} \widehat{\F}_{t\mid t-1} - \boldsymbol{\Lambda} \boldsymbol{\mu}_{z_t})     \\
			      \mathbf{P}_{t\mid t}             & =(\I-\K_t \boldsymbol{\Lambda}) \widehat{\mathbf{P}}_{t\mid t-1}.
		      \end{aligned}
	      \end{equation*}
	      The sampling steps are
	      \begin{itemize}
		      \item $\F_T \mid \tilde{\y}_T \sim N(\F_{T\mid T}, \mathbf{P}_{T\mid T})$, where
		            \begin{equation*}
			            \begin{aligned}
				            \F_{T\mid T} & =\widehat{\F}_{T \mid T-1}+\K_T(\y_{T}-\Lambda \widehat{\F}_{T\mid T-1} - \boldsymbol{\Lambda} \boldsymbol{\mu}_{z_T}) \\
				            \mathbf{P}_{T\mid T} & =(\I-\K_T \boldsymbol{\Lambda}) \widehat{\mathbf{P}}_{T\mid T-1}
			            \end{aligned}
		            \end{equation*}
		      \item $\F_t \mid \F_{t+1}, \tilde{\y}_t \sim N(\F_{t\mid t, \F_{t+1}}, \mathbf{P}_{t\mid t, \F_{t+1}})$, where
		            \begin{equation*}
			            \begin{aligned}
				            \F_{t\mid t,\F_{t+1}} & = \F_{t\mid t}+\mathbf{P}_{t\mid t} \A_{z_t}^T(\A_{z_t} \mathbf{P}_{t\mid t} \A_{z_t}^T+\HH_{z_{t+1}})^{-1}(\F_{t+1}-\A_{z_t} \F_{t\mid t}) \\
				            \mathbf{P}_{t\mid t,\F_{t+1}} & = \mathbf{P}_{t\mid t}-\mathbf{P}_{t\mid t} \A_{z_t}^{T}(\A_{z_t} \mathbf{P}_{t\mid t} \A_{z_t}^{T}+\HH_{z_{t+1}})^{-1} \A_{z_t} \mathbf{P}_{t\mid t}
			            \end{aligned}
		            \end{equation*}
	      \end{itemize}

	\item $\Q \mid \tilde{\F}_T, \tilde{\y}_T, \boldsymbol{\Omega}, \boldsymbol{\mu}_0, \boldsymbol{\mu}_1, \boldsymbol{\mu}_2, \boldsymbol{\Lambda}$.
	      Yield residual covariance matrix $\Q=\text{diag}(\sigma^2_1, \cdots, \sigma^2_N)$ is assumed to be diagonal. The update of each $\sigma_i^2$ term follows the standard inverse-Gamma conjugate sampling. For simplicity, we show the update of one regime, and multiple regimes proceed similarly.
	      \begin{equation*}
		      \sigma^2_i \mid \tilde{\F}_T, \tilde{\y}_T, \boldsymbol{\Omega}, \boldsymbol{\mu}_0, \boldsymbol{\mu}_1, \boldsymbol{\mu}_2, \boldsymbol{\Lambda} \sim IG\left(\alpha+\frac{T}{2}, \beta+\frac{1}{2} \sum_{t=1}^T (\y_{ti}-\boldsymbol{\Lambda}_i \F_t-\boldsymbol{\Lambda}_i \boldsymbol{\mu}_{z_t})^2\right),
	      \end{equation*}
	      where $\boldsymbol{\Lambda}_i$ is the $i$-th row in $\boldsymbol{\Lambda}$ and $\y_{ti}$ is the $i$-th row in $\y_t$, $i=1, 2, \cdots, N$.

	\item $\HH_g \mid \tilde{\F}_T, \A_0, \A_1, \A_2$.
	      Let $T_g$ be number of time periods in the regime $g$, and for $t$ satisfying the regime indicator $z_t=g$, $\eta_t \sim N(0, \HH_g)$. The update step is
	      \begin{equation*}
		      \HH_g\mid \tilde{\F}_T, \A_0, \A_1, \A_2, \boldsymbol{\mu}_0, \boldsymbol{\mu}_1, \boldsymbol{\mu}_2 \sim IW(\m_0+T_g, \M_0+\G_g),
	      \end{equation*}
	      where $\G_g=\sum\limits_{t: z_t=g} (\F_{t}-\A_{z_{t-1}}\F_{t-1})(\F_{t}-\A_{z_{t-1}}\F_{t-1})^T$.

	\item $\A_g \mid \boldsymbol{\gamma}_g, \tilde{\F}_T, \HH_0, \HH_1, \HH_2$. Let $\abf_g={\text{vec}}(\A_g^T)$, $\boldsymbol{\gamma}_g = {\text{vec}}((\boldsymbol{\gamma}^g)^T)$. 
		      {\scriptsize
			      \begin{equation*}
				      \begin{aligned}
					       & \pi(\abf_g\mid \tilde{\F}_T, \HH_0, \HH_1, \HH_2, \boldsymbol{\gamma}_g) \propto p(\F_{t+1:z_t=g} \mid \abf_g, \HH_0, \HH_1, \HH_2, \boldsymbol{\gamma}_g) \pi(\abf_g\mid \boldsymbol{\gamma}_g)                                                                    \\
					       & \propto \exp\left\{-\frac{1}{2}\left[\abf_g^T\left(\sum_{t:z_t=g} \HH_{z_{t+1}}^{-1}\otimes (\F_{t} \F_{t}^T)\right)\abf_g -
					      2\abf_g^T \left(\sum_{t:z_t=g} {\text{vec}}(\F_{t}\F_{t+1}^T \HH_{z_{t+1}}^{-1})\right)\right]\right\}                           \times \prod_{j,k=1}^3 \pi(\abf_{jk}^g\mid \gamma_{jk}^g)                         \\
					       & \propto \exp\left\{-\frac{1}{2}\left[\abf_g^T\left(\sum_{t:z_t=g} \HH_{z_{t+1}}^{-1}\otimes (\F_{t} \F_{t}^T)\right)\abf_g -
					      2\abf_g^T \left(\sum_{t:z_t=g} {\text{vec}}(\F_{t}\F_{t+1}^T \HH_{z_{t+1}}^{-1})\right)\right]\right\}                              \times \prod_{j,k=1}^3\exp\{-\frac{1}{2\xi_{\gamma_{jk}^g}^2}{a_{jk}^g}^2\} \\
					       & \propto \exp\left\{-\frac{1}{2}\left[\abf_g^T\left(\U_g^{-1}+\sum_{t:z_t=g} \HH_{z_{t+1}}^{-1}\otimes (\F_{t} \F_{t}^T)\right)\abf_g -
						      2\abf_g^T \left(\sum_{t:z_t=g} {\text{vec}}(\F_{t}\F_{t+1}^T \HH_{z_{t+1}}^{-1})\right)\right]\right\},
				      \end{aligned}
			      \end{equation*}
		      }where $\U_g=\text{diag}(\xi_{\gamma_{jk}^g}^2)$, $j, k=1,2,3$. Draw $\abf_g$ from conditional posterior $N(\bar{\abf}_g, \bar{\DD}_g)$, where
	      \begin{equation*}
		      \bar{\DD}_g = (\U_g^{-1}+\sum_{t:z_t=g} \HH_{z_{t+1}}^{-1}\otimes (\F_{t} \F_{t}^T))^{-1},\quad
		      \bar{\abf}_g = \bar{\DD}_g (\sum_{t:z_t=g} {\text{vec}}(\F_{t}\F_{t+1}^T \HH_{z_{t+1}}^{-1})).
	      \end{equation*}

	\item ${\gamma}_{jk}^g\mid \boldsymbol{\gamma}_{-jk}^g, \tilde{\F}_T, \HH_0, \HH_1, \HH_2$. $\boldsymbol{\gamma}_{-jk}^g$ represents the remaining elements except the $(j,k)$-th element of $\boldsymbol{\gamma}^g$.
		      {\small 
			      \begin{align*}
					       & \pi(\boldsymbol{\gamma}_g\mid \tilde{\F}_T, \HH_0, \HH_1, \HH_2) \propto \int p(\F_{t+1:z_t=g}\mid \abf_g, \HH_0, \HH_1, \HH_2, \boldsymbol{\gamma}_g) p(\abf_g\mid \boldsymbol{\gamma}_g)\pi(\boldsymbol{\gamma}_g) d \abf_g                                             \\
					       & \propto \int \exp\left\{-\frac{1}{2}\left[\abf_g^T\left(\sum_{t:z_t=g} \HH_{z_{t+1}}^{-1}\otimes (\F_{t} \F_{t}^T)\right)\abf_g -
					      2\abf_g^T \left(\sum_{t:z_t=g} {\text{vec}}(\F_{t}\F_{t+1}^T \HH_{z_{t+1}}^{-1})\right)\right]\right\}                                                                                                   \\
					       & \times \mid \U_g\mid^{-\frac{1}{2}} \exp\left(-\frac{1}{2}\abf_g^T \U_g^{-1} \abf_g\right) w^{\sum{{\gamma}_{jk}^g}} (1-w)^{\sum{(1-{\gamma}_{jk}^g)}} d \abf_g                                               \\
					       & = \mid \U_g\mid^{-\frac{1}{2}} w^{\sum{{\gamma}_{jk}^g}} (1-w)^{\sum{(1-{\gamma}_{jk}^g)}}                                                                                                          \\
					       & \times \int \exp\left\{-\frac{1}{2}\left[\abf_g^T\left(\U_g^{-1} + \sum_{t:z_t=g} \HH_{z_{t+1}}^{-1}\otimes (\F_{t} \F_{t}^T)\right)\abf_g -
					      2\abf_g^T \left(\sum_{t:z_t=g} {\text{vec}}(\F_{t}\F_{t+1}^T \HH_{z_{t+1}}^{-1})\right)\right]\right\}                                                                                                   \\
					       & \propto \mid \U_g\mid^{-\frac{1}{2}} \mid \bar{\DD}_g\mid^{\frac{1}{2}} \exp\left(\frac{1}{2} \bar{\abf}_g^T \bar{\DD}_g^{-1} \bar{\abf}_g\right) w^{\sum{\gamma_{jk}^g}} (1-w)^{\sum{(1-\gamma_{jk}^g)}} \\
					       & = \mid \U_g\mid^{-\frac{1}{2}} \mid \bar{\DD}_g\mid^{\frac{1}{2}} \exp\left(\frac{1}{2} \lbf^T \bar{\DD}_g \lbf\right) w^{\sum{\gamma_{jk}^g}} (1-w)^{\sum{(1-\gamma_{jk}^g)}},                           \\
				      \end{align*}
          }
	      where $\lbf=\sum_{t:z_t=g} {\text{vec}}(\F_{t}\F_{t+1}^T \HH_{z_{t+1}}^{-1})$. Therefore the conditional posterior of $\gamma_{jk}^g\mid \boldsymbol{\gamma}_{-jk}^g$ is
	      \begin{equation*}
		      \pi(\gamma_{jk}^g\mid \boldsymbol{\gamma}_{-jk}^g, \tilde{\F}_T, \HH_0, \HH_1, \HH_2) \propto \xi_{\gamma_{jk}^g}^{-1} \mid \bar{\D}_g\mid^{\frac{1}{2}} \exp\left(\frac{1}{2} \lbf^T \bar{\D}_g \lbf \right) w^{\sum{\gamma_{jk}^g}} (1-w)^{\sum{(1-\gamma_{jk}^g)}},
	      \end{equation*}
       which is a Bernoulli distribution. For diagonal elements of $\A_g$, force $\gamma_{jk}^g=1$. It's natural to assume that the dynamics of one factor is related to its historical values, so we don't impose spike-and-slab prior on diagonal elements of transition matrix $\A$ and let the corresponding $\gamma$ fixed at 1.

\end{itemize}
 
\begin{itemize}
	\item $\boldsymbol{\mu}_g \mid \tilde{\F}_T, \tilde{\y}_T, \boldsymbol{\Lambda}, \Q$.
	      Assume there are $T_g$ time points in regime $g$, and in regime $g$, the likelihood is
	      \begin{equation*}
		      \begin{aligned}
			       & p(\y_{t:z_t=g} \mid \bmu_g)  \propto \prod_{t:z_t=g} \exp\left\{-\frac{1}{2} \boldsymbol{\varepsilon}_{t}^T \Q^{-1} \boldsymbol{\varepsilon}_{t}\right\}                                                                             \\
			       & \propto \exp\left\{-\frac{1}{2}\left[\bmu_g^T\left(T_g \boldsymbol{\Lambda}^T \Q^{-1}\boldsymbol{\Lambda}\right)\bmu_g -  2\bmu_g^T \left(\sum_{t:z_t=g} (\boldsymbol{\Lambda}^T \Q^{-1}\y_{t}-\boldsymbol{\Lambda}^T \Q^{-1}\boldsymbol{\Lambda} \F_{t})\right)\right]\right\}.
		      \end{aligned}
	      \end{equation*}
	      The posterior distribution is
	      \begin{equation*}
		      \bmu_g \mid \tilde{\F}_T, \tilde{\y}_T, \boldsymbol{\Lambda}, \Q \sim N(\bar{\bmu}_g, \bar{\B}_g),
	      \end{equation*}
	      where
	      \begin{equation*}
		      \bar{\B}_g = ({\underline{\B}}^{-1}+T_g \boldsymbol{\Lambda}^T \Q^{-1}\boldsymbol{\Lambda})^{-1}
	      \end{equation*}
	      \begin{equation*}
		      \bar{\bmu}_g = \bar{\B}_g \left({\underline{\B}}^{-1} \underline{\bmu}+\sum_{t:z_t=g} (\boldsymbol{\Lambda}^T \Q^{-1}\y_{t}-\boldsymbol{\Lambda}^T \Q^{-1}\boldsymbol{\Lambda} \F_{t})\right).
	      \end{equation*}
\end{itemize}

\begin{itemize}
	\item $\lambda \mid \tilde{\F}_T, \bmu_0, \bmu_1, \bmu_2, \tilde{\y}_T, \Q $.
	      We use random walk Metropolis-Hastings algorithm to estimate $\lambda$. Superscript $t$ indicates the $t$-th update of MCMC.
	      \begin{itemize}
		      \item Generate $\lambda^{\ast}$ from proposal distribution $J_t(\lambda^{\ast} \mid \lambda^{(t-1)}) = U(0.01, 0.1)$.

		      \item Compute acceptance ratio $r$. $p_0(\lambda) = p(\lambda \mid \tilde{\F}_T, \bmu_0, \bmu_1, \bmu_2, \tilde{\y}_T, \Q)$ is the target posterior  distribution.
		            \begin{equation*}
			            r = \frac{p_0(\lambda^{\ast})}{p_0(\lambda^{(t-1)})} \times \frac{J_t(\lambda^{(t-1)} \mid \lambda^{\ast})}{J_t(\lambda^{\ast} \mid \lambda^{(t-1)})} = \frac{p_0(\lambda^{\ast})}{p_0(\lambda^{(t-1)})}
		            \end{equation*}

		      \item Sample $u \sim U(0,1)$, if $u < r$, set $\lambda^{t} = \lambda^{\ast}$, else set $\lambda^{t} = \lambda^{(t-1)}$.
	      \end{itemize}
	      The conditional posterior distribution is
		      {\footnotesize
			      \begin{equation*}
				      \begin{aligned}
					      p(\lambda \mid \tilde{\F}_T, &\bmu_0, \bmu_1, \bmu_2, \tilde{\y}_T, \Q)  \propto p(\tilde{\y}_T \mid \lambda, \tilde{\F}_T, \bmu_0, \bmu_1, \bmu_2, \Q) \times p(\lambda).                                                                                 \\
					                                                                       & = \prod_{t=1}^{T} p(\y_t \mid \bmu_0, \bmu_1, \bmu_2, \lambda) p(\lambda)                                                                                                  \\
					                                                                       & \propto \prod_{t=1}^{T} \frac{1}{\sqrt{2 \pi}} \mid \Q \mid^{-\frac{1}{2}} \exp\left\{-\frac{1}{2}\left(\y_t-\boldsymbol{\Lambda} (\F_t + \bmu_{z_t})\right)^T \Q^{-1} \left(\y_t-\boldsymbol{\Lambda} (\F_t + \bmu_{z_t})\right)\right\} \\
					                                                                       & \propto \exp\left\{- \frac{1}{2} \sum_{t=1}^T \left(\y_t - \boldsymbol{\Lambda} (\F_t + \bmu_{z_t})\right)^T \Q^{-1} \left(\y_t - \boldsymbol{\Lambda} (\F_t + \bmu_{z_t})\right)\right\}.
				      \end{aligned}
			      \end{equation*}
		      }

\end{itemize}

{\onehalfspacing
	\addcontentsline{toc}{section}{References}
	\bibliographystyle{chicago}
	\bibliography{ref.bib}  
}

\end{document}